\documentclass[preprint,12pt, authoryear]{elsarticle}
\usepackage[utf8]{inputenc}
\usepackage{geometry}
\usepackage{pdflscape}
\usepackage{afterpage}
\usepackage{tikzpagenodes} 	

\RequirePackage[pdftex,
      colorlinks={true},
      linkcolor={Blue3},
      urlcolor={Blue3},
      citecolor={Blue3},
      pdfstartview={FitH},
      bookmarks={true},
            pdfauthor={Author},
            pdftitle={Title},
            pdfsubject={Subject}
            plainpages=false,
            pdfpagelabels
            ]{hyperref}
\usepackage{graphicx}
\usepackage{xcolor}
\usepackage{booktabs}
\usepackage{subcaption}
\usepackage{adjustbox}
\usepackage{float}
\usepackage{amssymb}
\usepackage[font=small,skip=0pt]{caption}
\usepackage{multirow}
\usepackage{comment}
\usepackage{threeparttable}
\usepackage[symbol]{footmisc}
\definecolor{navyblue}{rgb}{0.0, 0.0, 0.5}

\usepackage{mathtools} 

\usepackage{enumitem}

\usepackage[autostyle]{csquotes}  

\usepackage{geometry}
\begin{document}
\title{Can transit investments in low-income neighbourhoods increase transit use? Exploring the nexus of income, car-ownership, and transit accessibility in Toronto}

\author[1]{Elnaz Yousefzadeh Barri$^*$}
\ead{barri16@itu.edu.tr}
\cortext[*]{Corresponding author}

\author[2,3]{Steven Farber}

\author[4]{Anna Kramer}

\author[5]{Hadi Jahanshahi}

\author[3]{Jeff Allen}

\author[1]{Eda Beyazit}

\address[1]{Department of Urban and Regional Planning, Istanbul Technical University, Turkey}
\address[2]{Department of Human Geography, University of Toronto Scarborough, Canada}
\address[3]{Department of Geography and Planning, University of Toronto,, Canada}
\address[4]{School of Urban Planning, McGill University, Canada}
\address[5]{Data Science Lab, Ryerson University, Canada}

\journal{Transportation Research Part D: Transport and Environment}

\begin{abstract}
Transportation equity advocates recommend improving public transit in low-income neighbourhoods to alleviate socio-spatial inequalities and increase quality of life. However, transportation planners often overlook transit investments in neighbourhoods with “transit-captive” populations because they are assumed to result in less mode-shifting, congestion relief, and environmental benefits, compared to investments that aim to attract choice riders in wealthier communities. In North American cities, while many low-income households are already transit users, some also own and use private vehicles. It suggests that transit improvements in low-income communities could indeed result in more transit use and less car use. Accordingly, the main objective of this article is to explore the statistical relationship between transit use and transit accessibility as well as how this varies by household income and vehicle ownership in the Greater Toronto and Hamilton Area (GTHA). Using stratified regression models, we find that low-income households with one or more cars per adult have the most elastic relationship between transit accessibility and transit use; they are more likely to be transit riders if transit improves. However, we confirm that in auto-centric areas with poor transit, the transit use of low-income households drops off sharply as car ownership increases. On the other hand, a sensitivity analysis suggests more opportunities for increasing transit ridership among car-deficit households when transit is improved. These findings indicate that improving transit in low-income inner suburbs, where most low-income car-owning households are living, would align social with environmental planning goals.
\end{abstract}


\begin{highlights}
    \item Transit use of low-income groups drops off quickly as car ownership increases.
    \item The elasticity of transit accessibility is the highest for low-income car-owners.
    \item Most ridership growth from accessibility is predicted for car-deficit households.
    \item Link to the paper: \url{https://doi.org/10.1016/j.trd.2021.102849}
\end{highlights}
\maketitle

\section{Introduction}
Historically, transit infrastructure investments have largely focussed on attracting choice riders in an effort to take cars off the road, and reap congestion and environmental benefits~\citep{Bhattacharjee2012, carey2002, pucher2002}. As a result, many socioeconomically disadvantaged communities, home to transit-dependent populations, were largely overlooked during the transit planning processes of the post-war era. The rationale being that investing in low-income neighbourhoods, where transit ridership is already very high, would be less likely to result in mode-shifting, congestion relief, and environmental benefits. More recently, justice and equity objectives for transportation investments are receiving growing attention in both research and planning practice. Focus is shifting towards the alleviation of transport disadvantages to encourage fairness in the opportunity for people to reach daily activity destinations. With the justice turn in transportation planning, we now know much more about the social benefits of achieving more equity in the distribution of transit benefits among population groups, including rationales grounded in theoretic~\citep{Lucas2012, Martens2016} and empirical work~\citep{Allen2020planning, stanley2011}. But in striving for equity, must we put aside our desires to similarly achieve the conventional benefits of congestion relief and environmental benefits? How true is the received wisdom that investments in “transit-dependent” communities will not result in sizable benefits associated with growth in transit mode share? 

In this paper, we argue that too little is known about the degree of transit demand in low-income communities, and how sensitive low-income populations are to transit accessibility improvements. More research is needed to better understand whether the social goals associated with low-income transit investments can align with the congestion and environmental goals associated with growing transit mode shares. Accordingly, this study argues for the importance of considering traditional goals in conjunction with those of social equity through travel behaviour change in low-income neighbourhoods. Using a large-scale travel survey in the Greater Toronto and Hamilton Area (GTHA), this study investigates how different income and car-ownership groups respond to transit accessibility improvements. Previous empirical studies have confirmed that transit investments in disadvantaged communities achieve social goals; they enhance social inclusion by unlocking supressed demand for trips and creating new opportunities to participate in daily activities~\citep{Allen2020planning, Lucas2009, Martens2016}. We further explore the extent to which transit investments in low-income neighbourhoods are likely to increase transit use, therefore reducing vehicle kilometres travelled (VKT), traffic congestion, air pollution, and other externalities. Consequently, the findings can guide planners and policy-makers to take account of transit investments in low-income neighbourhoods for alleviating both transport and financial burdens while reaping the societal benefits of positive environmental and congestion outcomes.

\section{Literature review}
Planners have traditionally focused on the value of transit investments and their efficiency in terms of environmental sustainability and value-of-time savings. This measurement regime ultimately supports the goal of reducing car-based trips via attracting choice riders to transit~\citep{Richmond2001}. From this perspective, numerous transport agencies and planners are evaluating the performance of rail projects with a congestion-relief target or conducting air quality analyses. For instance, \citet{Bhattacharjee2012} have analysed how successful the newly opened light rail system in Denver was in increasing transit ridership, taking cars off the road, and thus relieving congestion. They measured the temporal and spatial changes in the levels of highway traffic in terms of VMT changes. Their main purpose was investigating the spatial distribution of riders who switch from car to transit reflected in average VMT changes over 16 years. Research by \citet{Baum-Snow2005} explored the effects of new or extended transit rail between 1970 and 2000 on the transit mode share in sixteen major U.S. cities. Their study illustrated that rail transit projects do not necessarily lead to an overall increase in transit ridership; instead, increases in rail transit ridership stemmed from those switching from bus to rail. Nevertheless, these conventional approaches overlooked socioeconomically disadvantaged communities, their needs, and behaviour during the transit investment process.

From a political economy perspective, transportation projects and investments are theorized to be formed by the interest of powerful actors or power associations. These influential groups intervene in policy-making decisions to support their values and interests~\citep{Glaeser2018}. There may be evidence of this in the Toronto case, with much of the transit expansion in the region occurring in the form of commuter rail lines that mainly link middle- and upper-income suburban areas to the Central Business District. By focussing investments in suburban rail expansion for long-distance commuters, the needs of inner-suburban residents risk going unmet~\citep{Giuliano2005, brown2009}. Indeed, the City of Toronto has the majority of North America’s highest ridership bus routes, almost all operating in mixed traffic, and under crowded or crushed conditions during both peaks. These riders remain disempowered due to systemic marginalization along with income, race, and immigration lines~\citep{hertel2016, lo2011, Palm2020Social}. Moreover, they have largely been unsuccessful in attracting transit improvements for their daily, local travel needs.

Several travel surveys and studies have demonstrated that economically and socially disadvantaged groups, particularly low-income households, use public transit more frequently than other socioeconomic categories~\citep{Giuliano2005, pucher2003, rosenbloom1998}. Furthermore, due to structural racism and sexism, racialized people, women, and non-binary people are more likely to have lower incomes and are more likely to face safety issues when traveling from harassment and threat of violence~\citep{oswin2014, scholten2019}. Their greater reliance on transit partly implies why policymakers overlook them in the traditional transit planning process. Despite the higher level of reliance on transit for daily trips by low-income households, studies have shown that low-income households in the GTHA also make fewer and shorter trips~\citep{Allen2020planning, Paez2009}. This difference could be due to the high costs of trips and activity participation – whether in time or money, or due to other time-geographical or accessibility limitations. This suppressed demand offers an opportunity to improve mobility equity by removing barriers and equalizing the number of trips regardless of income, ceteris paribus. Achieving this goal is made difficult by the relatively recent reversal of the income-distance gradient observed across many global cities, including the GTHA~\citep{Kneebone2010, Glaeser2008}. Poverty is increasing in the suburbs partly due to inner-city gentrification and the changing geography of affordable housing~\citep{Ding2016, Ellen2011, pucher2003}. The combination of the auto centric design of cities with the suburbanization of poverty has resulted in a large group of financially constrained drivers who are driving because of a lack of alternatives, as well as transit users living in poorly served neighbourhoods far from social and economic activities. Consequently, these conditions serve to suppress activity participation, or shift travel burdens unduly on already structurally marginalized groups, further worsening the risks of social exclusion~\citep{Allen2021, Lucas2012, Martens2016}.

In recent years, car-dominant countries have witnessed the rapid increase in car ownership among low-income households~\citep{Blumenberg2014}. Focusing on the impact of private vehicles on trips, \citet{Blumenberg2012} have identified the profound impact of car ownership on the increase in travelled miles of low-income adults. Furthermore, some researchers have found that auto ownership plays a crucial role in accessing employment opportunities and higher earnings for vulnerable groups~\citep{Gurley2005, Raphael2002}. \citet{Baum2009}, for instance, measured the effect of car ownership on the probability of employment using a longitudinal survey, concluding that owning a car significantly produces positive employment outcomes and promotes welfare receipt exits. \citet{Curl2018} voiced a similar concern when evaluating the trend of car ownership regarding financial difficulties for households living in disadvantaged communities between 2006 and 2011 in Glasgow. They found that a large number of car owners keep their cars despite experiencing economic stresses because they consider it as a necessity for reaching their life opportunities. Their findings showed that having children and searching for a job deter the majority of low-income households in deprived neighbourhoods to relinquish their private vehicles. Moreover, the necessity of having a private vehicle to meet mobility needs forces them to buy a car, even if they are unwilling or cannot afford it~\citep{Curl2018, Potoglou2008, pucher2003}. Therefore, it may be true that transit investments in those low-income neighbourhoods will not help with mode shift because car-ownership brings a variety of opportunities for the poor, and they are committed to car-use after the heavy investment made in the car.

Other studies have shown unstable car ownership trends in low-income communities. This population segment more frequently changes its car ownership rate across time compared to other groups in society~\citep{klein2017}. Using a mobility biography approach, \citet{klein2019} have examined how a variety of life events affects the car ownership decisions of households over ten years. Their findings revealed that losing a job or worsening health has the most significant effect on giving up a car for the poor. Besides, looking at the financial burden of having a car, \citet{currie2007} explored the relationship between car-related expenditures (e.g., car purchase, insurance, and charges) and its financial difficulties for low-income car owners living on the fringe of Melbourne. They found that low-income, car-owning households living in outer Melbourne make fewer and shorter distance trips compared to other car owners in the same region. Interestingly, these fewer trips are highly reliant on their private vehicles and less frequently done by transit. They prefer car trips to transit trips, probably because using their vehicles provides a reduced cost of travel. Therefore, they do not opt to pay for transit when they can drive comfortably and less costly to their destinations. Consequently, they suggest transit investments could address transport disadvantage and mitigate financial burdens on vulnerable households as they believe that people in poverty will give up their private vehicles.

Concerning the above arguments, in this study, we explore whether transit investments in low-income neighbourhoods are likely to result in increased transit use, thus congestion and environmental co-benefits, as well as reducing socio-spatial inequalities for disadvantaged communities. To this end, we will investigate two contradictory arguments mentioned in the literature. First, if owning a private vehicle bears a substantial burden on low-income households, they are expected to display more sensitivity to transit accessibility improvements and to be more likely to switch their mode from car to transit. Second, if low-income households are either already transit users or reluctant to shed their car after their sizeable investment, then improving transit accessibility in low-income neighbourhoods will not necessarily be associated with mode shifting. Notably, the cost of auto ownership for a family includes expenses for car purchase, lease, loans, fuel, insurance, maintenance, parking, etc. However, our dataset does not include this information. Research shows that owning a private vehicle, including car-related expenditures, carries a significant cost to households in the lowest income quintile within the US, Canadian, and Australian contexts~\citep{Curl2018, currie2007, Deka2002, Walks2018}. The exact trade-offs between car-ownership costs and transit use cannot be explored in the current study due to data limitations.

\section{Methodology}
\subsection{Study context}
We conduct this study in a contemporary Canadian context, the Greater Toronto and Hamilton Area (GTHA). It is the largest urban agglomeration in Canada, with a population of more than 7 million people based on 2019 population estimates~\citep{Statistics2021} and one of the fastest-growing regions in North America. The GTHA contains the cities of Toronto and Hamilton and four regional municipalities including, Durham, York, Peel, and Halton. Metrolinx, the Greater Toronto Transportation Authority, is responsible for developing Regional Transportation Plans (RTP), operating commuter rail, and more recently, planning for all new heavy rail development (e.g., LRT, BRT, and subways) in the GTHA.  The overall public transportation system in the GTHA comprises nine local transit agencies, operating subway lines, surface buses, and streetcar routes, together with regional bus and rail lines. Toronto’s transit system operated by the Toronto Transit Commission (TTC) includes four subway lines, surface bus, and streetcar routes. A regional rail and commuter bus system, GO Transit (operated by Metrolinx), connects suburban regions to themselves and the city center, and is used primarily for long-distance commutes. Other municipalities in the region primarily offer local bus services, with some lite BRT functionality along select corridors. 

\begin{figure}[!ht]
    \centering
    \includegraphics[width=\textwidth]{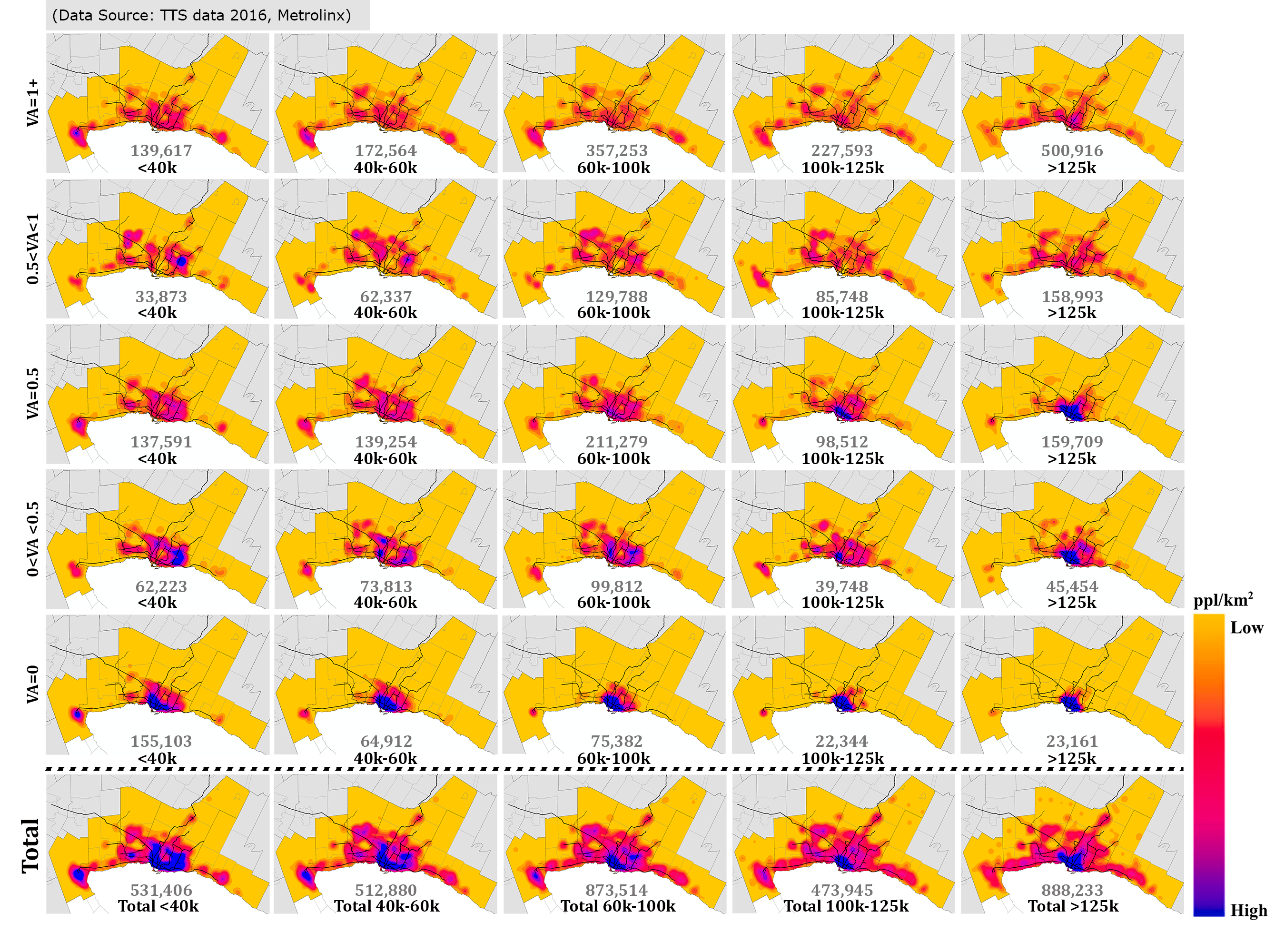}
    \caption{The population density of different household income and car-ownership classes (missing data are excluded). The number of people in each class is normalized by the total population of the same class and the total number of people in each census tract to eliminate the size effect. The color indicates the normalized persons per square kilometer.  VA = Vehicles per adult in the household}
    \label{fig:popdensity}
\end{figure}

Despite overall economic growth in Canada, its major cities have encountered growing socio-spatial inequalities, with increasing polarization as neighbourhoods change over time~\citep{ades2012, hulchanski2010}. Toronto is now the most unevenly distributed metro area in Canada, according to the Gini coefficient for income, and within the Toronto area, inequalities between neighbourhoods are very high~\citep{dinca2017}. Figure~\ref{fig:popdensity} displays the population density patterns of our study area for different income and car-ownership levels together with the population of each stratum. In the bottom row of Figure~\ref{fig:popdensity}, the ‘U’ shape pattern illustrates the higher concentrations of low-income households (<\$40k) in downtown Toronto and its inner suburbs. Regarding households with zero vehicles per adult (VA=0), as income increases, they become more concentrated in downtown Toronto, whereas the low-income carless cohort is more dispersed in the region.

Previous studies in the GTHA have shown that vulnerable groups have, on average, shorter transit travel times for their work commutes, and higher levels of accessibility than their counterparts~\citep{El-Geneidy2016Non-stop, Foth2013}. This can largely be explained by a) the vestiges of a sizable inner-city low-income population, and b) the proliferation of middle and upper-income households throughout the vast and poorly served outer suburbs of the region. Despite this overall distribution, recent work shows that there are hundreds of thousands of low-income households located in low-accessibility parts of the GTHA~\citep{Allen2019}. 

\begin{figure}[!ht]
    \centering
    \includegraphics[width=1\textwidth]{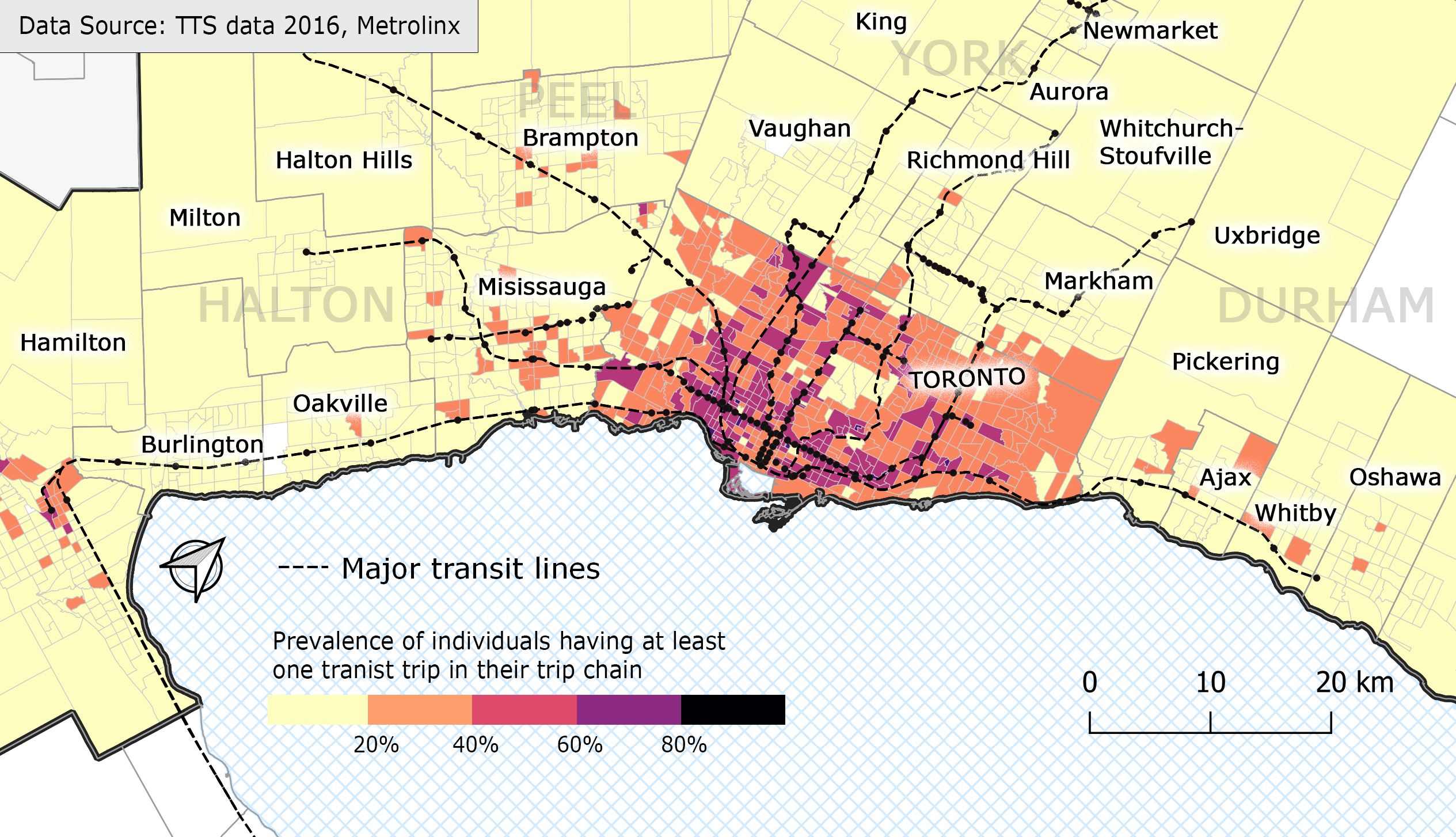}
    \caption{Percentage of individuals taking at least one transit trip in their daily trip chain }
    \label{fig:CurrentTransitUsers}
\end{figure}

Given the concentration of rapid transit within the City of Toronto, Figure~\ref{fig:CurrentTransitUsers} illustrates that most transit riders are living within the Toronto municipal boundaries, with concentrations closely mirroring both transit levels of service as well as the “U” shaped pattern of low-income brackets. In between the dense urban core and the poorly served outer suburbs, lies a transition zone characterized by people still having access to moderate levels of public transit, largely aligned with the service area of the TTC within the City of Toronto; consequently, 20-40\% of residents keep transit in their daily trip basket in this zone. Furthermore, there is a decline in transit use in the city centre, mainly due to the availability of biking and walking options for reaching destinations despite the high level of transit accessibility.

\subsection{Travel Survey Data }
This study is based on data from the 2016 Transportation Tomorrow Survey (TTS), a large-sample, one-day household travel diary. For the 2016 survey cycle, the targeted sampling rate was 5\% of households, except for Hamilton that only had a 3\% sampling rate target due to insufficient local funding. To provide a reliable population estimation, a set of expansion factors is included in the TTS 2016 data. The data expansion process consists of data weighting, considering dwelling type, household size, and the distribution of the population by age and gender matching the population distributions of the 2016 Canadian Census~\citep{TTS}. As researchers, we have access to the anonymized, individual and trip-level data. In this research, we only consider a subset of individual trips, those starting from and terminating in GTHA regions. Furthermore, we limit our investigation to adults aged 18 years or older –i.e., working-aged people assumed to be making autonomous residential location, vehicle ownership, and daily travel decisions.

\subsubsection{Descriptive analysis}
Table~\ref{tab:destriptive} provides a descriptive summary of our dataset, including a total of 122,724 households (249,632 individuals) aged 18 years and older and a total of 538,364 trips. These figures are expandable to 5,387,081 people, 2,532,632 households, and 11,610,043 trips, respectively.

\begin{table}[!ht]
    \centering
    \caption{Descriptive statistics of explanatory variables for respondents ($n=249,632$ ; expandable to $\mathcal{N}= 5,387,081$)}
    \label{tab:destriptive}
    \resizebox{0.55\textwidth}{!}{
    \begin{tabular}{lr|r}
    \toprule
    \multirow{2}{*}{\textbf{Variables}} & \multicolumn{2}{c}{\textbf{Individuals in the GTHA}} \\
    \cmidrule(lr){2-3}
     &\textbf{ Expanded frequency} & \textbf{Expanded Proportion} \\
     \midrule
    \textit{Age group} &  &  \\
    \qquad 18-25 & 702,598 & 13.04\% \\
    \qquad 26-35 & 964,695 & 17.91\% \\
    \qquad 36-45 & 955,252 & 17.73\% \\
    \qquad 46-55 & 1,057,659 & 19.63\% \\
    \qquad 56-65 & 841,574 & 15.62\% \\
    \qquad 65+ & 853,731 & 15.86\% \\
    \qquad Missing & 11,572 & 0.21\% \\
     &  &  \\
    \textit{Gender} &  &  \\
   \qquad  Female & 2,806,404 & 52.10\% \\
    \qquad Male & 2,580,677 & 47.90\% \\
     &  &  \\
    \textit{Household’s total income per year} &  &  \\
    \qquad \$0   to \$39,999 & 838,021 & 15.56\% \\
    \qquad \$40,000   to \$59,999 & 713,772 & 13.25\% \\
    \qquad \$60,000   to \$99,999 & 1,148,963 & 21.33\% \\
    \qquad \$100,000   to \$124,999 & 598,691 & 11.11\% \\
    \qquad \$125,000   and above & 1,089,156 & 20.22\% \\
    \qquad Missing & 998,478 & 18.53\% \\
     &  &  \\
    \textit{Vehicles per adult} &  &  \\
    \qquad (VA=0) & 597,833 & 11.10\% \\
    \qquad (0\textless{}VA\textless{}0.5) & 610,964 & 11.34\% \\
    \qquad (VA=0.5) & 1,243,951 & 23.10\% \\
    \qquad (0.5\textless{}VA\textless{}1) & 835,877 & 15.51\% \\
    \qquad (VA=1+) & 2,097,184 & 38.93\% \\
    \qquad Missing & 1,272 & 0.02\% \\
     &  &  \\
    \textit{Household size} &  &  \\
    \qquad One-person & 622,417 & 11.55\% \\
    \qquad Two-people & 1,419,702 & 26.35\% \\
    \qquad Three-people & 1,088,185 & 20.20\% \\
    \qquad Four-people & 1,199,803 & 22.27\% \\
    \qquad Five or more people & 1,056,974 & 19.62\% \\
     &  &  \\
    \textit{Employment status} &  &  \\
    \qquad Full time employment & 2,694,603 & 50.02\% \\
    \qquad Part time employment & 545,006 & 10.12\% \\
    \qquad Work at home (full time or part time) & 255,900 & 4.75\% \\
    \qquad Not employed (including students) & 1,888,816 & 35.06\% \\
    \qquad Missing & 2,756 & 0.05\% \\
     &  &  \\
    \textit{Possession of a driver’s license} &  &  \\
    \qquad Having driver’s license & 4,423,009 & 82.10\% \\
    \qquad Not having driver’s license & 858,401 & 15.93\% \\
    \qquad Missing & 105,671 & 1.96\% \\
     \midrule
     & \textbf{Mean} & \textbf{SD} \\
     \midrule
   \textit{ Population Density (per person)} & 6,864 & 7,924 \\
    \textit{Business Density (per person)} & 700 & 1,422 \\
    \textit{Intersection Density (per person)} & 54 & 36 \\
    \bottomrule
\end{tabular}
}
\end{table}

\begin{figure}[!ht]
    \centering
    \includegraphics[width=0.7\textwidth]{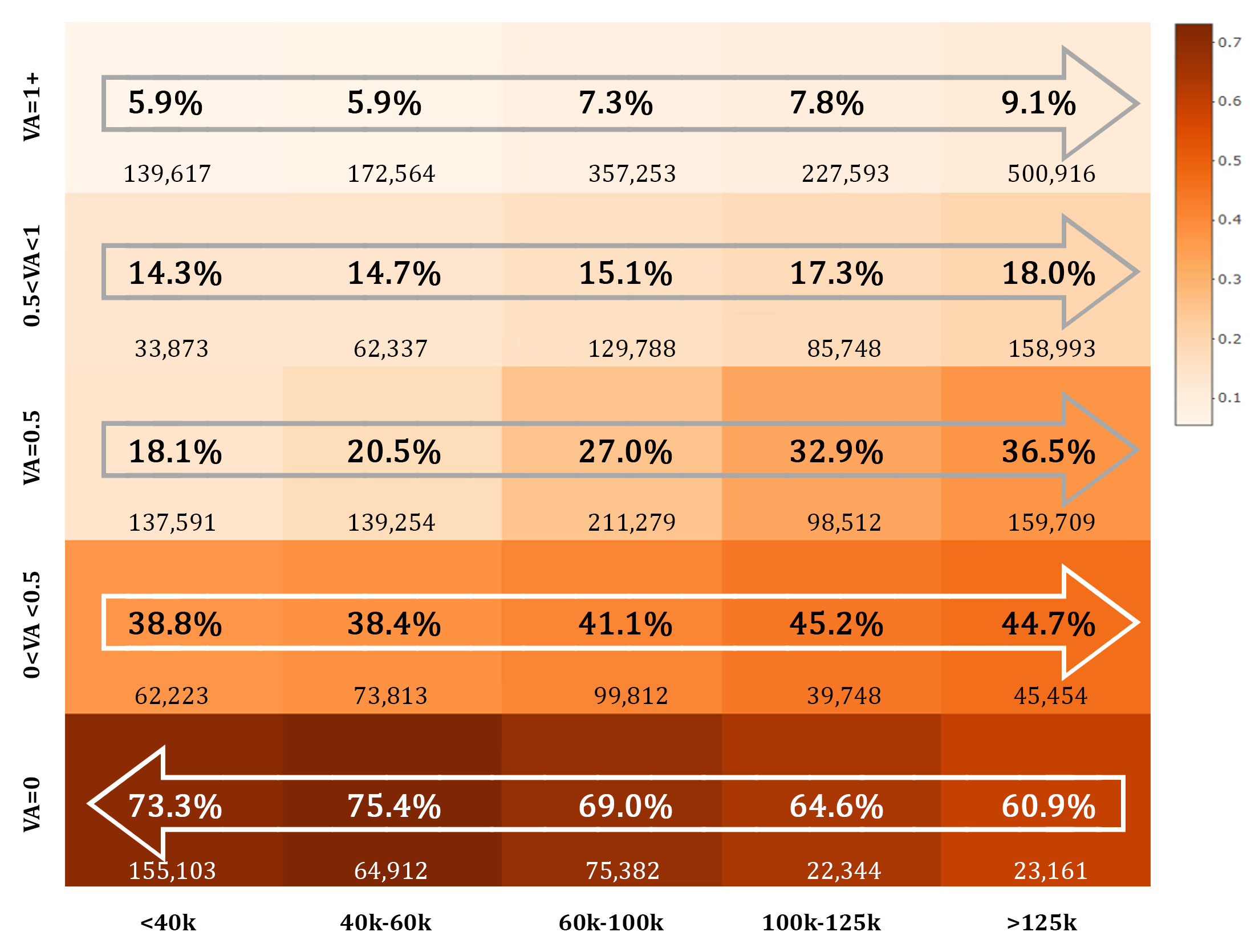}
    \caption{The percentage of individuals using transit in each class (people with missing income data and no trips are excluded).}
    \label{fig:Transit_users_precentage}
\end{figure}

Figure~\ref{fig:Transit_users_precentage} demonstrates a cross-tabulation of income, vehicle ownership, and the percentage of individuals with at least one transit trip per day. While carless households overall have very high rates of transit use, the rates are highest among low-income households (<\$40k) at 73.3\% vs. 60.9\% for the wealthier households (\$125k). Moreover, while there are more than 155,000 people living in carless low-income households, there are only 23,000 in carless wealthier households\footnote{It is important to note that the TTS income categories are designed to give granularity at the lower-end of the income scale, with nearly a full quarter of households earning higher than \$125,000.}. Notwithstanding the potential for wealthier carless households to travel by taxi and ridehailing more easily than low-income carless counterparts, the transit-use gap between incomes is also informed by the maps in Figure~\ref{fig:popdensity}, showing that high-income carless households are extremely concentrated in the core of the city, with added ability to either walk or bike. Conversely, low-income carless households are dispersed into the inner suburbs, where there are more barriers to active travel.

Interestingly, for car-owners, Figure~\ref{fig:Transit_users_precentage} shows that the percentage of people that use transit tends to increase with income. Conversely, among carless households, transit use tends to decline with income, likely related to residential concentration of the carless wealthy, and the relative affordability of taxis. We also observe a much steeper drop-off in transit use when moving from 0 cars to 0.5 cars per adult among the low-income households compared to the wealthy. At face value, these statistics support the hypothesis that when low-income households purchase a car, given its large expense relative to income, they do so with intentions to use it fully. A study by \citet{Giuliano2005} declares that the poor own a car since it is their only solution for household maintenance and income earning. Also, for many trips in the GTHA, the marginal costs by car are far cheaper than for transit, making the use of a car a cost-saving decision, except for trips heading to downtown Toronto along rapid transit corridor.

Next, we investigate how transit accessibility is distributed by income and car-ownership strata in the GTHA. Access to jobs is a commonly used measure of transit benefits and can be a crucial predictor of travel behaviour~\citep{Allen2020planning,Foth2013, Sanchez2004, Tyndall2017}. We use job accessibility as a proxy for overall transit benefits, which is defendable given the high degree of correlation between access to jobs via transit, and access to other daily destination types. Gravity-based accessibility to jobs by transit, calculated in a recent analysis in the GTHA~\citep{Allen2019}, is used in this study. This measure estimates the total number of reachable jobs from each Dissemination Area (origin), a census geographical unit with a population of 400 to 700 persons. The gravity-based accessibility is computed as

\begin{equation}
    {A}_{i} = \sum_{j=1}^J O_i f(t_{ij})
    \label{eq:transit_acc}
\end{equation}

where $A_i$ is the accessibility measure in zone $i$, $O_j$ is the count of jobs found in census tract j, $f(t_{ij})$ is the impedance function used to operationalize the diminishing attraction of jobs with travel time, and $t_{ij}$ is the travel time between $i$ and $j$ estimated with OpenTripPlanner using GTFS and OpenStreetMap data as inputs. This travel time includes walking time to and from stops, waiting time for the transit vehicle, in-vehicle travel time by transit, and transferring time. The impedance function is defined as 

\begin{align}
    {f}(t_{ij}) = \begin{cases}
    180 (90+ t_{ij})^{-1} &  t_{ij} < 90 \\
    0                     &  \text{otherwise}
    \end{cases}
    \label{eq:impedance_function}  
\end{align}

giving a weight of 0.5 to a 30-minute trip, roughly equal to the median duration commute trip in the GTHA across all modes. In this function, the maximum travel time value is limited to 90 minutes since very few people travel to jobs more than 90-minutes away~\citep{Allen2020measure}. 

\begin{figure}[!ht]
    \centering
    \includegraphics[width=.8\textwidth]{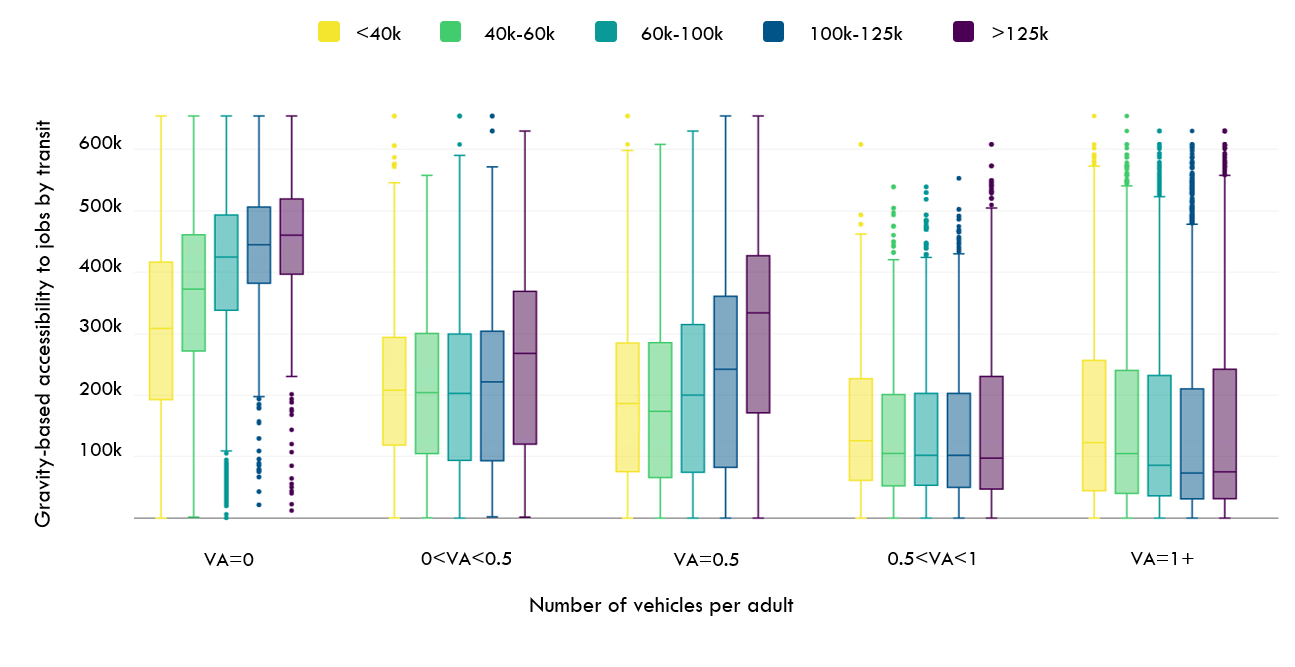}
    \caption{Distribution of accessibility, income and car ownership for all households.}
    \label{fig:boxplot_acc_income_car}
\end{figure}

Figure~\ref{fig:boxplot_acc_income_car} shows the relationship between accessibility, income, and car ownership levels in the study area. It illustrates that carless households, regardless of income level, tend to reside in neighbourhoods with higher levels of accessibility. Moreover, increasing income corresponds with an increase in transit accessibility for zero-car households. It suggests that high-income carless (and car-deficit, i.e., households with less than a car per driver) residents afford to locate in places with higher levels of transit accessibility. We also find that as the number of private vehicles increases, households tend to locate further from the core, in car-dependent neighbourhoods where transit accessibility levels are far lower, and the differences in accessibility are far less pronounced across income groups. The findings are consistent with the maps shown in Figure~\ref{fig:popdensity}. 

\subsection{Methods}
\subsubsection{Study approach}
The overall aim of this paper is to explore how transit investments, leading to changes in accessibility, affect transit use of different income and car ownership strata. We employ a zero-inflated negative binomial model to predict the number of transit trips that a person has per day. We implement our models in two phases: 1) using the whole sample to fit a comprehensive model on the population and 2) exploring 25 stratified models for different income and car-ownership levels (5 car ownership levels $\times$ 5 income levels). This allows us to explore the variability in response to accessibility. Following model estimation, we perform a sensitivity analysis to explore how changes to accessibility will differentially affect transit trip generation throughout the region. This provides an additional layer of policy-relevant analysis, enabling us to directly evaluate the potential for transit investments in low-income communities to unlock suppressed demand for transit travel.

\subsubsection{Model specification}
We estimate a zero-inflated negative binomial (ZINB) model to investigate the influence and significance of socio-demographic characteristics, local environments, and trip factors on the number of transit trips per individual. In our analysis, the dependant variable is the count of daily transit trips of an individual. We define a pool of candidate attributes that may affect taking transit after removing highly correlated covariates. The ZINB was selected because a large number of individuals in our data were without any transit trips in their travel day (n=116,451). ZINB models, unlike negative binomial, can deal with excess zeros and over-dispersion of the data~\citep{Sultana2018}. In particular, the model comprises two distinct processes: one for generating zero values with the probability $p_{i}$, and the other for generating counts from negative binomial with the probability $1-p_i$. The zero-inflation portion of the model consists of a binary logit model predicting non-occurrence, i.e., not taking transit, whereas the count portion of the model predicts the frequency of occurrence, i.e., the number of public transit trips. Therefore, the expected number of transit trips is computed as

\begin{equation}
    E(y_i) = p_i \times 0 + (1-p_i) \times n_i
    \label{eq:ZINB}
\end{equation}
where $n_i$ is the expected transit trip count given it is not zero. 

We use weighted ZINB in which the weights are normalized TTS expansion factors of each individual. Instead of using the direct expansion factors as the weight, we rescale the weights in a way that it sums to the stratum sample size. We normalize them by the mean of expansion factors per stratum. These weights correct biases that may occur due to non-representative sampling in the study region. Finally, we present the results of the zero-inflation portion and count portion of ZINB in the form of odds ratios and incidence rate ratios, respectively. They are obtained by exponentiating the coefficients of each of the model portions.

\subsection{ZINB model}
Table~\ref{tab:ZINB_model} contains the odds ratios (ORs), incidence rate ratios (IRRs), and significance levels of model coefficients. OR values, obtained from the zero-inflation portion, demonstrate the probability of having zero transit trips. In other words, the values greater than one shows the increase in the probability of not taking transit, and vice versa. For instance, the coefficient for transit accessibility shows a negative relationship for the zero-inflation portion and a positive relationship for the count portion of the model, indicating an increase in accessibility leads to an increase in the probability of taking transit trips\footnote{We also examined alternative definitions of the transit accessibility variable to account for nonlinearities (quadratic, cubic and sigmoid transformations), but they do not result in improved model fits or any changes in interpretation. Therefore, we consider only the linear effect of accessibility in our models to reduce their complexity.}. For car ownership, owning more vehicles per adult in a household reduces the likelihood of using transit. Unsurprisingly, households with one or more cars per adult have a significantly negative coefficient, indicating a lower probability of taking transit than the other car owners. This interpretation may also represent the association between poor transit service and a resident’s propensity to own a car. We explore this assumption in Section~\ref{sec:sensitivity_analysis}. Contrasting the effect size of car-ownership and income levels, we observe that the number of vehicles per adult has a much higher coefficient than income in the zero-inflation portion. Moreover, individuals with a driver’s licence, after controlling for car ownership, are more reluctant to take transit for their daily trips than those without a license. Likewise, free parking spots at the workplace reduce the likelihood of using transit. On the other hand, the coefficient of holding a transit pass is a significant predictor for taking transit. Males show 37.6 percent less inclination to take transit compared to females. Younger individuals have more propensity to take transit (becoming one-year older reduces the probability of taking transit by 2.8 percent).

\begin{table}[!ht]
    \centering
    \caption{ZINB model results (N= 3,279,979; n = 149,177)}
    \label{tab:ZINB_model}
    \resizebox{0.73\textwidth}{!}{
    \begin{tabular}{lll}
    \toprule
    \multicolumn{3}{l}{\textbf{Dep Var.= The number of Transit Trips}} \\
    \midrule
    \multirow{1}{*}{\textbf{Description of Independent Variables}} & \multicolumn{2}{c}{\textbf{ZINB Model}} \\
    \cmidrule(lr){2-3}
    & \multicolumn{1}{l}{\textbf{\begin{tabular}[c]{@{}l@{}}Probability of no \\ transit trip (OR)\textsuperscript{a,b}\end{tabular}}} & \multicolumn{1}{l}{\textbf{\begin{tabular}[c]{@{}l@{}}Incidence rate ratio \\ (IRR) for transit use\textsuperscript{a,b}\end{tabular}}} \\
    \midrule
    \textit{(intercept)} & 
    {0.007 ***} &  {1.070 *} \\
    Distance of mandatory trips (km) & 
    {0.989 ***} &  {1.004 ***} \\
    Distance of discretionary trips (km) & 
    {1.017 ***} & {1.005 ***} \\
    Age & 
    {1.029 ***} & {1.002 ***} \\
    
    \begin{tabular}[c]{@{}l@{}}Household’s total income per year \\ 
    \textit{(ref. category: \textless \$40k)}\end{tabular} & {} & {} \\
    \qquad \$40k-\$60k & {1.112 *} & {0.987} \\
    \qquad \$60k-\$100k & {1.027} & {0.932 ***} \\
    \qquad \$100k-\$125k & {1.036} & {0.920 ***} \\
    \qquad \$125k+ & {1.138 **} & {0.919 ***} \\
    \begin{tabular}[c]{@{}l@{}}Number of vehicles per adult\\   
    \textit{(ref. category: VA=0)}\end{tabular} &  &  \\
    \qquad 0 \textless VA \textless 0.5 & 
    {11.414   ***} & {0.948 ***} \\
    \qquad VA=0.5 & 
    {24.769  ***} & {0.921 ***} \\
    \qquad 0.5 \textless VA \textless 1 & 
    {31.107  ***} & {0.901 ***} \\
    \qquad VA=1+ & 
    {78.198  ***} & {0.877 ***} \\
    \multirow{2}{*}{\begin{tabular}[c]{@{}l@{}}Gender\\
    \textit{(ref. category: Female)}\end{tabular}} & {} & {} \\
     & {1.371 ***} & {0.986} \\
    \begin{tabular}[c]{@{}l@{}}Free parking at workplace\\ 
    \textit{(ref. category: No)}\end{tabular} & {} & {} \\
    \qquad Yes & {11.691  ***} & {0.980} \\
    \qquad NA & {4.839 ***} & {1.014} \\
    \multirow{2}{*}{\begin{tabular}[c]{@{}l@{}}Having driving license\\ 
   \textit{ (ref. category: No)}\end{tabular}} & {} & {} \\
     & {7.477 ***} & {0.970 **} \\
    \multirow{2}{*}{\begin{tabular}[c]{@{}l@{}}Having transit pass\\ 
    \textit{(ref. category: No)}\end{tabular}} & {} & {} \\
     & {0.058 ***} & {1.414 ***} \\
    Measure of accessibility to jobs using a gravity \\
    function (transit commute) & {0.572 ***} &  {1.069 ***} \\
    \begin{tabular}[c]{@{}l@{}} Population density \textsubscript{c} \end{tabular} & 
    {1.057 **} & {1.004} \\
    \begin{tabular}[c]{@{}l@{}} Business density \textsubscript{c} \end{tabular} & 
    {1.056 ***} & {0.953 ***} \\
    \begin{tabular}[c]{@{}l@{}} Intersection density \textsubscript{c} \end{tabular} & 
    {1.002 ***} & {0.999 ***} \\
    \bottomrule
    \multicolumn{3}{l}{\begin{tabular}[c]{@{}l@{}} {\footnotesize \textsuperscript{a} Significance codes: *** $p < 0.001$, ** $p < 0.01$, * $p < 0.05$} \end{tabular}} \\
    \multicolumn{3}{l}{\begin{tabular}[c]{@{}l@{}} {\footnotesize \textsuperscript{b} People with no trips or an extraordinary number of trips greater than 25 are removed.} \end{tabular}} \\
    \multicolumn{3}{l}{\begin{tabular}[c]{@{}l@{}} {\footnotesize \textsuperscript{c} Local built environment characteristics of travelers come from the weighted sum of values normalized by area}\\ {\footnotesize in each Dissemination Area.} \end{tabular}} \\
    \end{tabular}
}
\end{table}

Since travel mode choice is a function of the built environment~\citep{Cervero1997}, we append residential neighbourhood characteristics for each individual to the dataset. Accordingly, intersection density as a design metric comes from the total number of 3-way or more intersections per square kilometer. The population density in each Dissemination Area is from the 2016 Canadian Census, and business density comes from the Canadian business registry. These variables are measured as the sum of individuals and businesses per square kilometer, respectively. After controlling for transit accessibility, the coefficients for population and business density show a negative association with using transit. Similarly, the intersection density is negatively associated with using transit. We are somewhat puzzled by these results, but assume that they indicate that higher densities are associated with high levels of active travel, something that is not discernable within a single-mode model like the ZINB. Having long discretionary trips is a deterrent to using transit; however, longer mandatory trips have a positive association with the probability of taking transit. 

On the other hand, the count-model IRR values greater than one show a positive impact on taking more transit trips, and those less than one have a negative impact on the number of transit trips. Considering the count model’s IRR, free parking spot at workplace, population density and gender are not a significant predictor of the overall number of transit trips. Moving from the reference low-income group to households with total income greater than \$125k per year corresponds to a \%12.3 decline in having more transit trips. Moreover, having more vehicles per adult in a household decreases the number of transit trips. Similarly, individuals with a driver’s license have a 2.3\% lower transit trip rate. Conversely, holding a transit pass increase the number of transit trips by 41.4\%. Both the longer discretionary and mandatory trips have a positive association with using more transit trips in their daily trips.

\subsubsection{Sensitivity analysis} \label{sec:sensitivity_analysis}
Next, we generate 25 stratified logistic regression models, one for each combination of income and car-ownership strata ($5 \times 5$). The objective of these models is to contrast the effect size of accessibility across different groups. As we want to see the probability of taking transit, we utilize logistic regression model only for this section. According to previous studies, the coefficient of the zero portion of the ZINB may be difficult to interpret by having structural and sampling zeros~\citep{Staub2013, hua2014}. Therefore, we switch to logistic regression for this task. To compare the effect size, we use an elasticity metric. Elasticity, as a unit-free measurement, is the ratio of the percentage change in an independent variable associated with the percentage change in a dependent variable. The elasticity of accessibility for each observation is defined as

\begin{align}
    E_{x_i} = \beta_i \times x_i(1-P_i)&& \forall x_i \in \mathbb{R}
    \label{eq:elasticity}
\end{align}
where $E_{x_i}$ is the elasticity of individual $i$, $\beta_i$ is the logistic regression coefficient of transit accessibility, $x_i$ is the transit accessibility value of individual $i$, and $P_i$ is the estimated probability of taking transit~\citep{train2009}. Then, we averaged these elasticities over the population in each stratum~\citep{Ewing2010}. 

\begin{align}
    \bar{E}_x=\frac{\sum_{i=1}^n w_i E_i}{\sum_{i=1}^n w_i}
    \label{eq:weighted_elasticity}
\end{align}
$\bar{E}_x$ is the weighted elasticity where $w_i$ is the expansion value of individual $i$.

\begin{figure}[!ht]
    \centering
    \includegraphics[width=.7\textwidth]{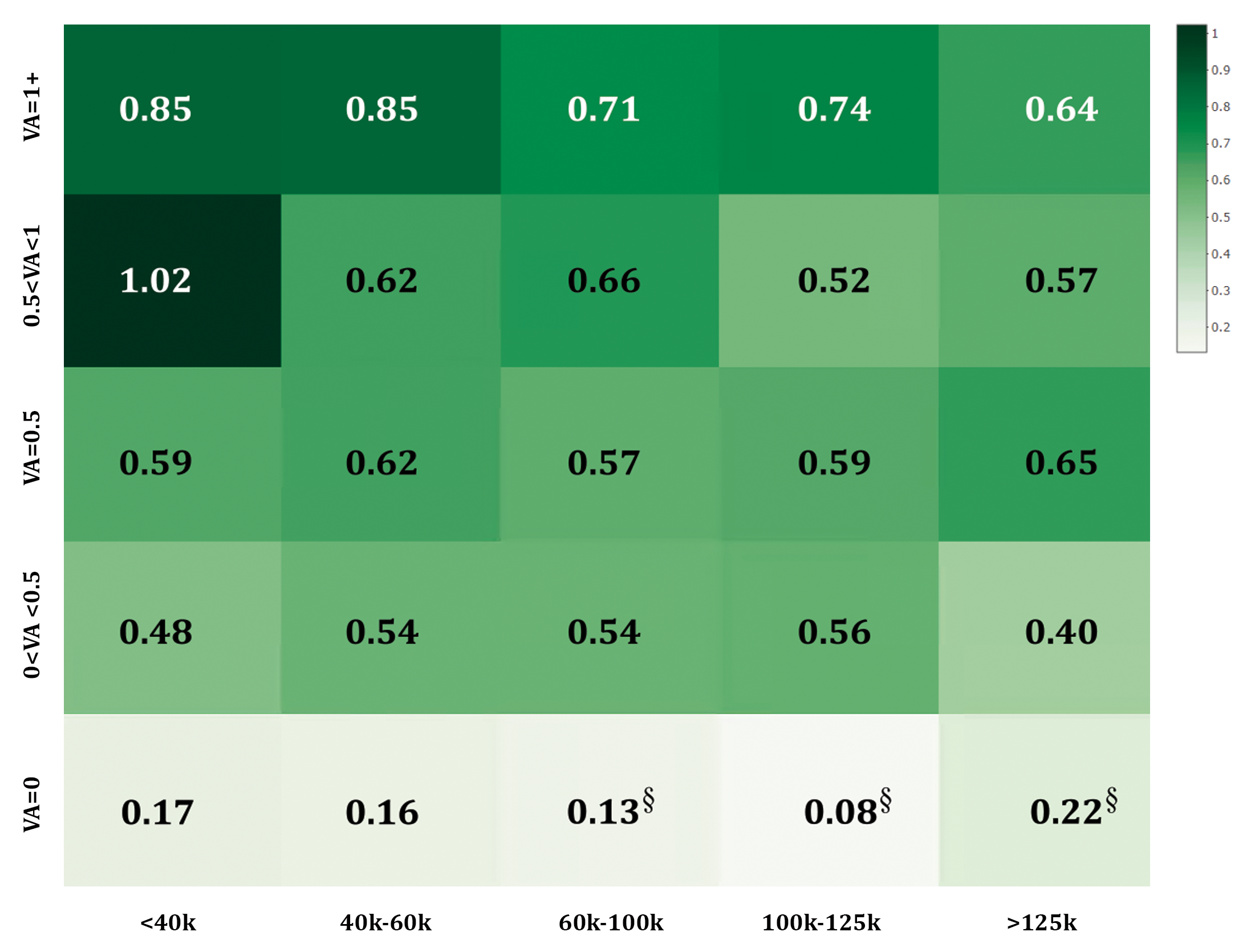}
    \caption{Elasticity estimates of transit accessibility for 25 logistic regression models (income and car ownership levels) \\
    $^\S$\textit{Elasticity estimates were insignificant at the 0.05 level.}}
    \label{fig:heatmap_elasticity_estimates}
\end{figure}

In Figure~\ref{fig:heatmap_elasticity_estimates}, each grid defines the elasticity of accessibility for the corresponding stratified logistic regression model. The weights of the model are the expansion factor of each stratum normalized by the mean expansion factor of the same group. Interpretation of elasticities are straightforward. For instance, the elasticity of 0.87 for the low-income households with one or more vehicles per adult indicates that a 1 percent increase in accessibility will result in a 0.87 percent increase in the probability of taking transit. Incidentally, low-income and high car-owning households appear to have the highest overall sensitivities to transit accessibility, indicating a latent demand for mode-switching if only transit were more readily provided to them. There are about 140,000 individuals in this strata, representing a large proportion (26\%) of the low-income population. Overall, the top row shows that all households with one or more vehicles per adult, readily take transit as transit accessibility improves. On the other hand, carless households in the high and medium-income category prove to be insensitive to the change in their accessibility, having a p-value less than 0.05. It can be the upshot of the fact that most carless wealthy households are already transit users, or live in places that allow for active travel lifestyles. Therefore, their transit accessibility cannot be further improved to increase the probability that they will use transit. Of course, improving accessibility may have other personal benefits for those carless households, such as less travel and waiting times, greater reliability, and less crowding. Interestingly, the effect of accessibility on transit ridership increases as households own more personal vehicles. People owning cars are optional transit riders, thus, enhancing their accessibility probably provides impetus to use public transport. The elasticities tell us which individuals are more or less sensitive to accessibility improvements. We next apply these elasticities to the GTHA’s population to ascertain how much opportunity there is to generate additional transit trips by focussing investments at different strata.

\begin{figure}[!ht]
    \centering
    \includegraphics[width=.9\textwidth]{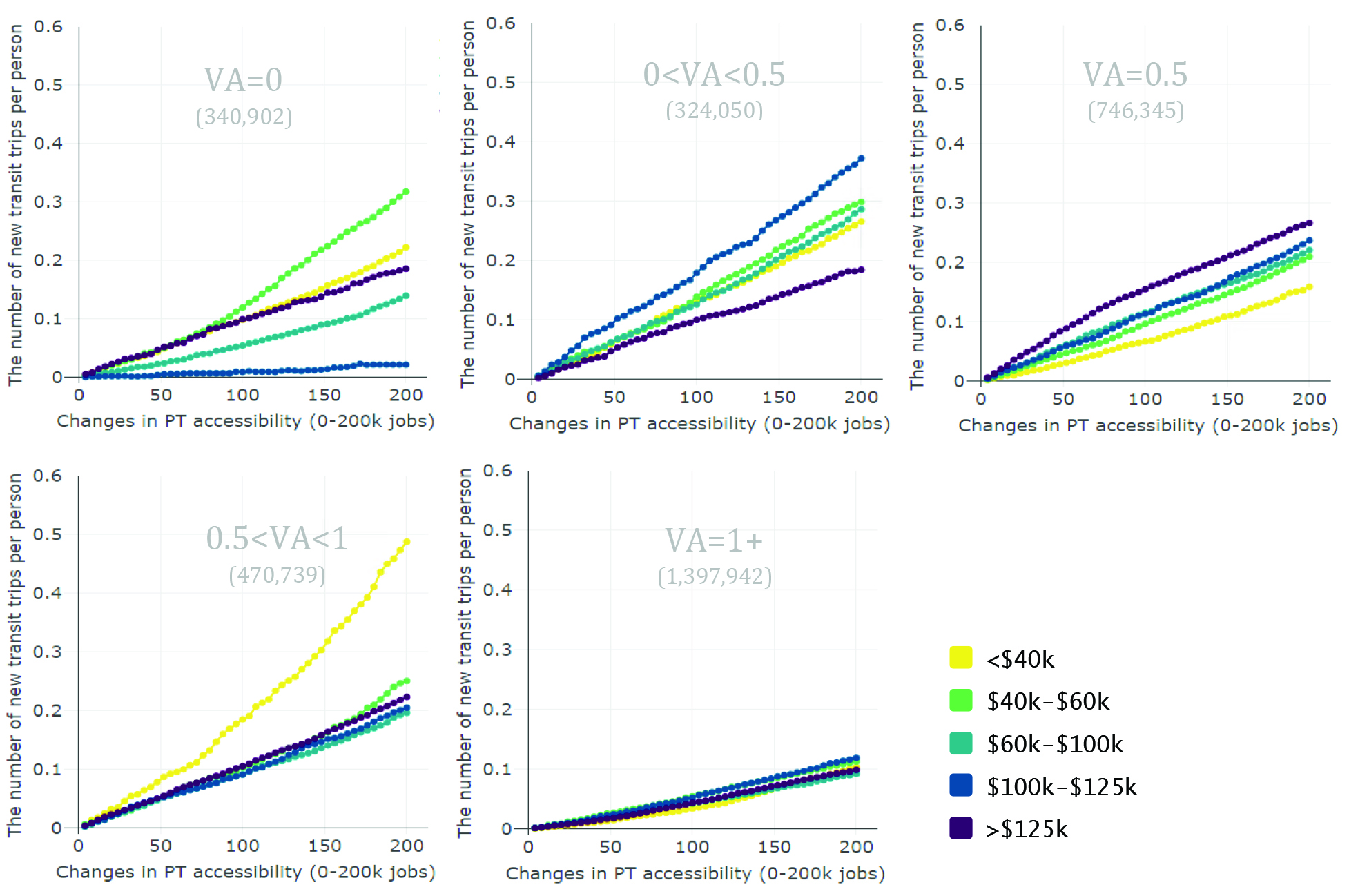}
    \caption{The expanded changes in transit trips per person by accessibility improvements. (The expanded population of each stratum is shown on the graphs)}
    \label{fig:expanded_changes}
\end{figure}

To estimate the number of new transit trips induced by a hypothetical transit investment, i.e., accessibility increase, we utilize the stratified ZINB models to determine the current number of transit trips ($y_i^c$) for individual i in class c as a baseline. Then, while all other independent variables remain constant, we incrementally increase the level of accessibility from 0 to 200k new jobs and estimate the new number of transit trips ($\hat{y}_i^c$) for each person. The smaller gains in accessibility (less than 50k jobs) would roughly be achievable by moderate investments in the existing transit system (e.g., more frequent service), whereas the largest accessibility gain (i.e., 200k jobs) requires significant improvements in transit infrastructure (e.g., new rapid transit)~\citep{Allen2020planning, Farber2017}. Having the baseline values, we compute the change in the number of transit trips per individual in our weighted sample as follows:

\begin{align}
    \Delta\hat{y}^c=\frac{\sum_{i=1}^{n^c}{\hat{y}_i^c} - \sum_{i=1}^{n^c}{y_i^c}}{n^c} && \forall c \in \{1 \dots 25\}
\end{align}
where $n^c$ is the total number of individuals in class $c$, and $\Delta{\hat{y}^c}$ is the predicted change in the number of transit trips due to a change in accessibility within population class $c$. The result of this analysis is demonstrated in Figure~\ref{fig:expanded_changes}. The y-axis shows the expanded number of newly generated daily transit trips per person in the GTHA by increasing transit accessibility across the GTHA. These numbers include both transit trips shifted from other modes and entirely new transit trips, and don’t differentiate between the two.

Noting a large number of carless households reside in places with a high level of accessibility, we still observe a significant discrepancy in sensitivities of various income groups to accessibility improvements. Figure~\ref{fig:expanded_changes} shows that among zero-car groups, more transit trips are induced among low-income groups. Notably, these low-income carless households already take more transit trips than other income and car-ownership brackets (Figure~\ref{fig:Transit_users_precentage}). Moreover, Figure~\ref{fig:heatmap_elasticity_estimates} showed the elasticity of non-transit riders of each class given accessibility improvement (converting from 0 to 1), while Figure~\ref{fig:expanded_changes} depicts the total number of newly generated transit trips per person —i.e., both new transit riders and increased transit trips among existing transit riders— after accessibility improvement. Figure~\ref{fig:newly_generated_tranist_trips} in
\ref{sec:new_transit_trip} shows how much transit ridership growth is associated with each source. Overall, households with one or more cars per person, regardless of their income level, are less responsive to accessibility increase than other car-ownership brackets. We may conclude that when households own more cars, they are more willing to use them even if transit accessibility was improved. We should also note that the accessibility coefficients of high and medium-income groups were insignificant for the carless strata in the models, meaning that these curves should only be used illustratively. Unsurprisingly, car-deficit households show more tendency toward taking transit after accessibility improvements because it opens a new door for them to select another travel mode. Notably, these households with less than a car per adult strongly incline to choose public transit since they have to share a car in a household.

Table~\ref{tab:expanded_new_transit_trips} shows the response of each income and car-ownership level to a hypothetical level of transit improvement. It indicates the estimated number of newly generated trips after improving accessibility equally, by 50,000 or 200,000 jobs, or relatively, by 10 or 25 percent above existing levels for each respondent. The changes in accessibility have the most disproportional impact on the three classes of households with less than one car per adult (0<VA<1). The results illustrate that in total, 64\% of new transit trips belong to these three groups although they are only 47\% of the whole population. They are mostly living in inner-suburbs with a medium level of accessibility (Figure~\ref{fig:boxplot_acc_income_car}) and sharing one car in a household. Therefore, accessibility improvements help these family members to have another mode option for reaching their destinations. This finding is consistent with a recent study by~\citet{blumenberg2020}. They also found that car-deficit households (i.e., 0<VA<1) are more likely to use public transit. On the contrary, transit riders with zero cars, on average, make only 13.2\% of the newly generated trips while comprising 10\% of the population. As a result, transit investments will contribute more to increasing transit trips of car-deficit households. Moreover, families who own more than one vehicle per adult and live in suburban car-dependent neighbourhoods are less likely to shift their travel mode. It is also notable that there is no significant difference in the relative increase in new transit trips for different income groups. The percentage of new transit trips for different income strata has the same distribution as their population. It strongly suggests that increases in transit ridership resulting from improvements in transit accessibility come from car deficit households.

\begin{table}[!ht]
    \centering
    \caption{The expanded number of new transit trips generated for each class after transit accessibility improvement}
    \label{tab:expanded_new_transit_trips}
    \resizebox{0.7\textwidth}{!}{
    \begin{tabular}{lr|rrrr}
    \toprule
    \multirow{2}{*}{\textbf{Strata}} & \multirow{2}{*}{\textbf{Population}} & \multicolumn{4}{c}{\textbf{Accessibility   Improvement}} \\
     &  & 
     \multicolumn{1}{c}{\textbf{50k}} & 
     \multicolumn{1}{c}{\textbf{200k}} & 
     \multicolumn{1}{c}{\textbf{10\%}} & 
     \multicolumn{1}{c}{\textbf{25\%}} \\
     \midrule
    \textbf{VA=0} & 340,902   (10\%) & 12,611 (10\%) & 70,417 (12\%) & 9,344 (15\%) & 25,855 (16\%) \\
    \textbf{0\textless{}VA\textless{}0.5} & 324,050   (10\%) & 20,532 (16\%) & 91,225 (16\%) & 9,134 (15\%) & 24,154 (15\%) \\
    \textbf{VA=0.5} & 746,345   (23\%) & 41,103 (33\%) & 163,889 (28\%) & 21,713 (35\%) & 54,365 (33\%) \\
    \textbf{0.5\textless{}VA\textless{}1} & 470,739   (14\%) & 24,265 (19\%) & 110,860(19\%) & 8,452 (13\%) & 21,002 (13\%) \\
    \textbf{VA=1+} & 1,397,942   (43\%) & 26,633 (21\%) & 142,837(25\%) & 13,998 (22\%) & 36,963 (23\%) \\
    \midrule
    \textbf{\textless{}\$40k} & 531,406 (16\%) & 19,044 (15\%) & 104,732 (18\%) & 10,155 (16\%) & 28,538 (18\%) \\
    \textbf{\$40k-\$60k} & 512,880   (16\%) & 21,229 (17\%) & 107,115 (18\%) & 10,490 (17\%) & 27,720 (17\%) \\
    \textbf{\$60k-\$100k} & 873,514   (27\%) & 33,181 (27\%) & 144,396 (25\%) & 16,031 (26\%) & 39,979 (25\%) \\
    \textbf{\$100k-\$125k} & 473,945   (14\%) & 18,519 (15\%) & 82,548 (14\%) & 7,764 (12\%) & 20,128 (12\%) \\
    \textbf{\textgreater{}\$125k} & 888,233   (27\%) & 33,171 (27\%) & 140,437 (24\%) & 18,201 (29\%) & 45,974 (28\%) \\
    \bottomrule
    \end{tabular}
}
\end{table}

\section{Conclusion}
The purpose of this study was to understand how increasing transit accessibility might differentially affect transit use of different income and car-ownership groups. We are particularly interested in whether transit investments in lower socio-economic neighbourhoods enhance transit mode share, either through increasing transit trips among existing transit riders or generating new transit riders. We hypothesized that low-income carless households may be largely insensitive to transit improvements, the so-called “captive transit” users for whom transit investments may not result in large environment or congestion co-benefits. The case for low-income car-owning households is less predictable, with competing arguments suggesting either: 

\setlength{\parindent}{20pt}  a) these households will be more sensitive to accessibility improvements since the costs of car ownership and use are high, and many could benefit from using transit rather than car if service levels were improved; or

b)	these households will be insensitive to accessibility improvements since once owning a car, it usually provides a reduced marginal cost of travel, and households with limited financial resources will not opt to pay for transit if they can drive places for “free” (or rather at a low marginal cost).

\noindent In response to this dichotomy, we found that transit use is more sensitive to transit investment in households that own cars, and most sensitive in low income, car-owning households (with an elasticity almost equal to 1). It is strong evidence in support of (a) that for non-transit rider low-income households, owning a car is a financial burden, and increased transit provides increased opportunity for mobility and transit use. At the same time, however, we found that the tendency for individuals in low-income households to take transit reduces dramatically as soon as their household owns a private vehicle, whereas it drops more gradually for high-income households. This is evidence in support of (b), indicating that low-income car-owning households can become extremely car reliant, and even less multi-modal than their wealthier car-owning comparators.

In both cases, the findings are supported by spatial analysis of where different population strata live vis a vis existing levels of transit supply or relative environmental dominance of the automobile. Wealthier carless households tend to concentrate in neighbourhoods where existing transit accessibility levels are high. This pattern becomes complicated as income levels decrease. Low-income, carless households, are more dispersed in lower-accessibility areas compared to higher-income, carless households, who are more concentrated in the very core of the city. 

Finally, since there is evidence to support that both (a) and (b) are true, the simulation analysis can provide some insight into how these forces combine to result in an overall transit ridership response in the region.  Our simulations, however crude, apply accessibility gains to individuals across the region to determine among which population groups we see the largest increase in transit use. Here, the evidence is quite clear; more new trips are predicted to be made by households with less than one car per adult. There may indeed be more opportunity for increasing transit use overall by targeting car-deficit households. Conversely, accessibility improvements in areas where the accessibility gap between transit and car is large, and a significant number of car-deficit low-income households are residing, would be an effective way of both increasing transit ridership and improving equity. Nonetheless, this does not preclude the possibility of first investing in targeted areas where low-income, car-owning, and transit sensitive populations are presently residing. In Figure~\ref{fig:popdensity}, we show that this population is mostly living in Toronto’s inner suburbs, indicating that both social and mode-shifting goals can be achieved if we make investments there.

Alternatively, we also provide evidence that housing policies in coordination with transportation policies are essential to facilitate the transit accessibility of poor households~\citep{pucher2003, Kramer2018}. Housing policies that provide affordable housing in areas with higher levels of accessibility will be a strategy to prevent low-income households from incurring financial burdens of car ownership. Future studies should investigate other models to validate the result of the research in different urban contexts. On the other hand, built environmental variables such as mixed land uses, walkable street networks, dense neighbourhoods, safety in neighbourhoods, employment, retail densities, and so on influence transit ridership. We suggest planners considering neighbourhood characteristics in evaluating local scale projects. Our findings reveal several important factors often ignored by policy-makers (e.g., the impact of transit investments on different car-ownership and income strata). Unlike the previous studies, we do not entirely rely on the model’s coefficients; instead, sensitivity analysis takes into consideration both the amount and the weight of transit accessibility for different strata.

Based on the result obtained, we believe more focussed research on the willingness of low-income car-owning families to switch to transit is needed. Future research can consider the trip chaining of individuals and pinpoint for which trip type (e.g., for home-to-school trips), the tendency of using public transit might be higher. Therefore, investigating the possible transit investments concerning the destinations and sociodemographic attributes of individuals may help planners develop more nuanced policies and investments. 

It is worth mentioning that the cross-sectional design of our study limits our ability to validate our findings over time. The snapshot of car-ownership and income level does not demonstrate whether low-income car-owning households are likely to give up their cars going forward. A longitudinal study in which periodic changes in car ownership can be measured would strengthen the analysis. Longitudinal analysis, gives a broader picture of the household’s changing decisions over time, making causal analysis more feasible, and allowing us to better estimate long-run behavioural responses to accessibility improvements.

Another noteworthy caveat is that long-term residential selection for the neighbourhoods with an improved transit system should be explored. There is a possibility that low-income households will obtain the ability to live car-free in those neighbourhoods after new transit investments. Perhaps the benefits are reaped via long-term shifts in residential preference and car-ownership decisions and not in the “momentary’’ shifts in behavior among people existing in their current accessibility and car ownership levels. Alternatively, we might find that over time the gains made by low-income residents are lost due to gentrification and displacement processes. Again, this shortcoming of the present study points towards the need for longitudinal analysis. 

In this paper, we studied the effects of accessibility improvement on transit use in the GTHA. As such, the findings of our study may not generalize to other regions. However, this area includes one of the most diverse cities in the world by having people from different backgrounds. Hence, the behaviour of the people is representative of many other similar contexts. Nonetheless, the goal of this paper is not to construct a theory that applies to all cities, but rather to shed light on the need to explore how transit investments would differentially impact people based on their incomes and car-ownership levels, and that these differences offer policy makers with a new angle to design transit plans that simultaneously achieve social equity goals and traditional transport planning goals that stem from inducing mode-shifting from car to transit.

\bibliographystyle{elsarticle-harv}
\bibliography{bib}

\begin{thebibliography}{58}
\expandafter\ifx\csname natexlab\endcsname\relax\def\natexlab#1{#1}\fi
\providecommand{\url}[1]{\texttt{#1}}
\providecommand{\href}[2]{#2}
\providecommand{\path}[1]{#1}
\providecommand{\DOIprefix}{doi:}
\providecommand{\ArXivprefix}{arXiv:}
\providecommand{\URLprefix}{URL: }
\providecommand{\Pubmedprefix}{pmid:}
\providecommand{\doi}[1]{\href{http://dx.doi.org/#1}{\path{#1}}}
\providecommand{\Pubmed}[1]{\href{pmid:#1}{\path{#1}}}
\providecommand{\bibinfo}[2]{#2}
\ifx\xfnm\relax \def\xfnm[#1]{\unskip,\space#1}\fi
\bibitem[{Ades et~al.(2012)Ades, Apparicio and S{\'e}guin}]{ades2012}
\bibinfo{author}{Ades, J.}, \bibinfo{author}{Apparicio, P.},
  \bibinfo{author}{S{\'e}guin, A.M.}, \bibinfo{year}{2012}.
\newblock \bibinfo{title}{Are new patterns of low-income distribution emerging
  in canadian metropolitan areas?}
\newblock \bibinfo{journal}{The Canadian Geographer/le g{\'e}ographe canadien}
  \bibinfo{volume}{56}, \bibinfo{pages}{339--361}.
\bibitem[{Allen and Farber(2019)}]{Allen2019}
\bibinfo{author}{Allen, J.}, \bibinfo{author}{Farber, S.},
  \bibinfo{year}{2019}.
\newblock \bibinfo{title}{Sizing up transport poverty: A national scale
  accounting of low-income households suffering from inaccessibility in canada,
  and what to do about it}.
\newblock \bibinfo{journal}{Transport Policy} \bibinfo{volume}{74},
  \bibinfo{pages}{214--223}.
\newblock \DOIprefix\doi{https://doi.org/10.1016/j.tranpol.2018.11.018}.
\bibitem[{Allen and Farber(2020a)}]{Allen2020measure}
\bibinfo{author}{Allen, J.}, \bibinfo{author}{Farber, S.},
  \bibinfo{year}{2020}a.
\newblock \bibinfo{title}{A measure of competitive access to destinations for
  comparing across multiple study regions}.
\newblock \bibinfo{journal}{Geographical analysis} \bibinfo{volume}{52},
  \bibinfo{pages}{69--86}.
\bibitem[{Allen and Farber(2020b)}]{Allen2020planning}
\bibinfo{author}{Allen, J.}, \bibinfo{author}{Farber, S.},
  \bibinfo{year}{2020}b.
\newblock \bibinfo{title}{Planning transport for social inclusion: An
  accessibility-activity participation approach}.
\newblock \bibinfo{journal}{Transportation Research Part D: Transport and
  Environment} \bibinfo{volume}{78}, \bibinfo{pages}{102212}.
\newblock \DOIprefix\doi{https://doi.org/10.1016/j.trd.2019.102212}.
\bibitem[{Allen and Farber(2021)}]{Allen2021}
\bibinfo{author}{Allen, J.}, \bibinfo{author}{Farber, S.},
  \bibinfo{year}{2021}.
\newblock \bibinfo{title}{Suburbanization of transport poverty}.
\newblock \bibinfo{journal}{Annals of the American Association of Geographers}
  \bibinfo{volume}{111}, \bibinfo{pages}{1833--1850}.
\newblock \DOIprefix\doi{10.1080/24694452.2020.1859981}.
\bibitem[{Baum(2009)}]{Baum2009}
\bibinfo{author}{Baum, C.L.}, \bibinfo{year}{2009}.
\newblock \bibinfo{title}{The effects of vehicle ownership on employment}.
\newblock \bibinfo{journal}{Journal of Urban Economics} \bibinfo{volume}{66},
  \bibinfo{pages}{151--163}.
\newblock \DOIprefix\doi{https://doi.org/10.1016/j.jue.2009.06.003}.
\bibitem[{Baum-Snow et~al.(2005)Baum-Snow, Kahn and Voith}]{Baum-Snow2005}
\bibinfo{author}{Baum-Snow, N.}, \bibinfo{author}{Kahn, M.E.},
  \bibinfo{author}{Voith, R.}, \bibinfo{year}{2005}.
\newblock \bibinfo{title}{Effects of urban rail transit expansions: Evidence
  from sixteen cities, 1970-2000 [with comment]}.
\newblock \bibinfo{journal}{Brookings-Wharton Papers on Urban Affairs} ,
  \bibinfo{pages}{147--206}\URLprefix
  \url{http://www.jstor.org/stable/25067419}.
\bibitem[{Bhattacharjee and Goetz(2012)}]{Bhattacharjee2012}
\bibinfo{author}{Bhattacharjee, S.}, \bibinfo{author}{Goetz, A.R.},
  \bibinfo{year}{2012}.
\newblock \bibinfo{title}{Impact of light rail on traffic congestion in
  denver}.
\newblock \bibinfo{journal}{Journal of Transport Geography}
  \bibinfo{volume}{22}, \bibinfo{pages}{262--270}.
\newblock \DOIprefix\doi{https://doi.org/10.1016/j.jtrangeo.2012.01.008}.
  \bibinfo{note}{special Section on Rail Transit Systems and High Speed Rail}.
\bibitem[{Blumenberg et~al.(2020)Blumenberg, Brown and
  Schouten}]{blumenberg2020}
\bibinfo{author}{Blumenberg, E.}, \bibinfo{author}{Brown, A.},
  \bibinfo{author}{Schouten, A.}, \bibinfo{year}{2020}.
\newblock \bibinfo{title}{Car-deficit households: determinants and implications
  for household travel in the us}.
\newblock \bibinfo{journal}{Transportation} \bibinfo{volume}{47},
  \bibinfo{pages}{1103--1125}.
\bibitem[{Blumenberg and Pierce(2012)}]{Blumenberg2012}
\bibinfo{author}{Blumenberg, E.}, \bibinfo{author}{Pierce, G.},
  \bibinfo{year}{2012}.
\newblock \bibinfo{title}{Automobile ownership and travel by the poor: Evidence
  from the 2009 national household travel survey}.
\newblock \bibinfo{journal}{Transportation Research Record}
  \bibinfo{volume}{2320}, \bibinfo{pages}{28--36}.
\newblock \DOIprefix\doi{10.3141/2320-04}.
\bibitem[{Blumenberg and Thomas(2014)}]{Blumenberg2014}
\bibinfo{author}{Blumenberg, E.}, \bibinfo{author}{Thomas, T.},
  \bibinfo{year}{2014}.
\newblock \bibinfo{title}{Travel behavior of the poor after welfare reform}.
\newblock \bibinfo{journal}{Transportation Research Record}
  \bibinfo{volume}{2452}, \bibinfo{pages}{53--61}.
\newblock \DOIprefix\doi{10.3141/2452-07}.
\bibitem[{Brown and Thompson(2009)}]{brown2009}
\bibinfo{author}{Brown, J.R.}, \bibinfo{author}{Thompson, G.L.},
  \bibinfo{year}{2009}.
\newblock \bibinfo{title}{The influence of service planning decisions on rail
  transit success or failure, mti report 08-04} .
\bibitem[{Carey(2002)}]{carey2002}
\bibinfo{author}{Carey, G.N.}, \bibinfo{year}{2002}.
\newblock \bibinfo{title}{Applicability of bus rapid transit to corridors with
  intermediate levels of transit demand}.
\newblock \bibinfo{journal}{Journal of public Transportation}
  \bibinfo{volume}{5}, \bibinfo{pages}{5}.
\bibitem[{Cervero and Kockelman(1997)}]{Cervero1997}
\bibinfo{author}{Cervero, R.}, \bibinfo{author}{Kockelman, K.},
  \bibinfo{year}{1997}.
\newblock \bibinfo{title}{Travel demand and the 3ds: Density, diversity, and
  design}.
\newblock \bibinfo{journal}{Transportation Research Part D: Transport and
  Environment} \bibinfo{volume}{2}, \bibinfo{pages}{199--219}.
\newblock \DOIprefix\doi{https://doi.org/10.1016/S1361-9209(97)00009-6}.
\bibitem[{Curl et~al.(2018)Curl, Clark and Kearns}]{Curl2018}
\bibinfo{author}{Curl, A.}, \bibinfo{author}{Clark, J.},
  \bibinfo{author}{Kearns, A.}, \bibinfo{year}{2018}.
\newblock \bibinfo{title}{Household car adoption and financial distress in
  deprived urban communities: A case of forced car ownership?}
\newblock \bibinfo{journal}{Transport Policy} \bibinfo{volume}{65},
  \bibinfo{pages}{61--71}.
\newblock \DOIprefix\doi{https://doi.org/10.1016/j.tranpol.2017.01.002}.
  \bibinfo{note}{household transport costs, economic stress and energy
  vulnerability}.
\bibitem[{Currie and Senbergs(2007)}]{currie2007}
\bibinfo{author}{Currie, G.}, \bibinfo{author}{Senbergs, Z.},
  \bibinfo{year}{2007}.
\newblock \bibinfo{title}{Exploring forced car ownership in metropolitan
  melbourne} .
\bibitem[{{Data Management Group}(2017)}]{TTS}
\bibinfo{author}{{Data Management Group}}, \bibinfo{year}{2017}.
\newblock \bibinfo{title}{Tts introduction}.
\newblock
  \bibinfo{howpublished}{\url{http://dmg.utoronto.ca/transportation-tomorrow-survey/tts-introduction}}.
\newblock \bibinfo{note}{Accessed: 21.06.2021}.
\bibitem[{Deka(2002)}]{Deka2002}
\bibinfo{author}{Deka, D.}, \bibinfo{year}{2002}.
\newblock \bibinfo{title}{Transit availability and automobile ownership: Some
  policy implications}.
\newblock \bibinfo{journal}{Journal of Planning Education and Research}
  \bibinfo{volume}{21}, \bibinfo{pages}{285--300}.
\newblock \DOIprefix\doi{10.1177/0739456X0202100306}.
\bibitem[{Dinca-Panaitescu et~al.(2017)Dinca-Panaitescu, Hulchanski, Laflèche,
  McDonough, Maaranen and Procyk}]{dinca2017}
\bibinfo{author}{Dinca-Panaitescu, M.}, \bibinfo{author}{Hulchanski, D.},
  \bibinfo{author}{Laflèche, M.}, \bibinfo{author}{McDonough, L.},
  \bibinfo{author}{Maaranen, R.}, \bibinfo{author}{Procyk, S.},
  \bibinfo{year}{2017}.
\newblock \bibinfo{title}{The Opportunity Equation in the Greater Toronto Area:
  An update on neighbourhood income inequality and polarization}.
\newblock \bibinfo{publisher}{United Way Toronto and York Region}.
\bibitem[{Ding et~al.(2016)Ding, Hwang and Divringi}]{Ding2016}
\bibinfo{author}{Ding, L.}, \bibinfo{author}{Hwang, J.},
  \bibinfo{author}{Divringi, E.}, \bibinfo{year}{2016}.
\newblock \bibinfo{title}{Gentrification and residential mobility in
  philadelphia}.
\newblock \bibinfo{journal}{Regional Science and Urban Economics}
  \bibinfo{volume}{61}, \bibinfo{pages}{38--51}.
\newblock \URLprefix
  \url{https://www.sciencedirect.com/science/article/pii/S0166046216301223},
  \DOIprefix\doi{https://doi.org/10.1016/j.regsciurbeco.2016.09.004}.
\bibitem[{El-Geneidy et~al.(2016)El-Geneidy, Buliung, Diab, van Lierop,
  Langlois and Legrain}]{El-Geneidy2016Non-stop}
\bibinfo{author}{El-Geneidy, A.}, \bibinfo{author}{Buliung, R.},
  \bibinfo{author}{Diab, E.}, \bibinfo{author}{van Lierop, D.},
  \bibinfo{author}{Langlois, M.}, \bibinfo{author}{Legrain, A.},
  \bibinfo{year}{2016}.
\newblock \bibinfo{title}{Non-stop equity: Assessing daily intersections
  between transit accessibility and social disparity across the greater toronto
  and hamilton area (gtha)}.
\newblock \bibinfo{journal}{Environment and Planning B: Planning and Design}
  \bibinfo{volume}{43}, \bibinfo{pages}{540--560}.
\newblock \DOIprefix\doi{10.1177/0265813515617659}.
\bibitem[{Elizabeth and Emily(2010)}]{Kneebone2010}
\bibinfo{author}{Elizabeth, K.}, \bibinfo{author}{Emily, G.},
  \bibinfo{year}{2010}.
\newblock \bibinfo{title}{The suburbanization of poverty: Trends in
  metropolitan america, 2000 to 2008}.
\newblock \bibinfo{journal}{Metropolitan Policy Program at Brookings} .
\bibitem[{Ellen and O'Regan(2011)}]{Ellen2011}
\bibinfo{author}{Ellen, I.G.}, \bibinfo{author}{O'Regan, K.M.},
  \bibinfo{year}{2011}.
\newblock \bibinfo{title}{How low income neighborhoods change: Entry, exit, and
  enhancement}.
\newblock \bibinfo{journal}{Regional Science and Urban Economics}
  \bibinfo{volume}{41}, \bibinfo{pages}{89--97}.
\newblock \DOIprefix\doi{https://doi.org/10.1016/j.regsciurbeco.2010.12.005}.
\bibitem[{Ewing and Cervero(2010)}]{Ewing2010}
\bibinfo{author}{Ewing, R.}, \bibinfo{author}{Cervero, R.},
  \bibinfo{year}{2010}.
\newblock \bibinfo{title}{Travel and the built environment}.
\newblock \bibinfo{journal}{Journal of the American Planning Association}
  \bibinfo{volume}{76}, \bibinfo{pages}{265--294}.
\newblock \DOIprefix\doi{10.1080/01944361003766766}.
\bibitem[{Farber and Marino(2017)}]{Farber2017}
\bibinfo{author}{Farber, S.}, \bibinfo{author}{Marino, M.G.},
  \bibinfo{year}{2017}.
\newblock \bibinfo{title}{Transit accessibility, land development and
  socioeconomic priority: A typology of planned station catchment areas in the
  greater toronto and hamilton area}.
\newblock \bibinfo{journal}{Journal of Transport and Land Use}
  \bibinfo{volume}{10}, \bibinfo{pages}{879--902}.
\bibitem[{Foth et~al.(2013)Foth, Manaugh and El-Geneidy}]{Foth2013}
\bibinfo{author}{Foth, N.}, \bibinfo{author}{Manaugh, K.},
  \bibinfo{author}{El-Geneidy, A.M.}, \bibinfo{year}{2013}.
\newblock \bibinfo{title}{Towards equitable transit: examining transit
  accessibility and social need in toronto, canada, 1996–2006}.
\newblock \bibinfo{journal}{Journal of Transport Geography}
  \bibinfo{volume}{29}, \bibinfo{pages}{1--10}.
\newblock \DOIprefix\doi{https://doi.org/10.1016/j.jtrangeo.2012.12.008}.
\bibitem[{Giuliano(2005)}]{Giuliano2005}
\bibinfo{author}{Giuliano, G.}, \bibinfo{year}{2005}.
\newblock \bibinfo{title}{Low income, public transit, and mobility}.
\newblock \bibinfo{journal}{Transportation Research Record}
  \bibinfo{volume}{1927}, \bibinfo{pages}{63--70}.
\newblock \DOIprefix\doi{10.1177/0361198105192700108}.
\bibitem[{Glaeser et~al.(2008)Glaeser, Kahn and Rappaport}]{Glaeser2008}
\bibinfo{author}{Glaeser, E.L.}, \bibinfo{author}{Kahn, M.E.},
  \bibinfo{author}{Rappaport, J.}, \bibinfo{year}{2008}.
\newblock \bibinfo{title}{Why do the poor live in cities? the role of public
  transportation}.
\newblock \bibinfo{journal}{Journal of Urban Economics} \bibinfo{volume}{63},
  \bibinfo{pages}{1--24}.
\newblock \DOIprefix\doi{https://doi.org/10.1016/j.jue.2006.12.004}.
\bibitem[{Glaeser and Ponzetto(2018)}]{Glaeser2018}
\bibinfo{author}{Glaeser, E.L.}, \bibinfo{author}{Ponzetto, G.A.},
  \bibinfo{year}{2018}.
\newblock \bibinfo{title}{The political economy of transportation investment}.
\newblock \bibinfo{journal}{Economics of Transportation} \bibinfo{volume}{13},
  \bibinfo{pages}{4--26}.
\newblock \DOIprefix\doi{https://doi.org/10.1016/j.ecotra.2017.08.001}.
  \bibinfo{note}{the political economy of transport decisions}.
\bibitem[{Gurley and Bruce(2005)}]{Gurley2005}
\bibinfo{author}{Gurley, T.}, \bibinfo{author}{Bruce, D.},
  \bibinfo{year}{2005}.
\newblock \bibinfo{title}{The effects of car access on employment outcomes for
  welfare recipients}.
\newblock \bibinfo{journal}{Journal of Urban Economics} \bibinfo{volume}{58},
  \bibinfo{pages}{250--272}.
\newblock \DOIprefix\doi{https://doi.org/10.1016/j.jue.2005.05.002}.
\bibitem[{Hertel et~al.(2016)Hertel, Keil and Collens}]{hertel2016}
\bibinfo{author}{Hertel, S.}, \bibinfo{author}{Keil, R.},
  \bibinfo{author}{Collens, M.}, \bibinfo{year}{2016}.
\newblock \bibinfo{title}{Next stop equity: Routes to fairer transit access in
  the greater toronto and hamilton area}.
\newblock \bibinfo{journal}{Toronto, ON} .
\bibitem[{Hua et~al.(2014)Hua, Wan, Wenjuan and Paul}]{hua2014}
\bibinfo{author}{Hua, H.}, \bibinfo{author}{Wan, T.}, \bibinfo{author}{Wenjuan,
  W.}, \bibinfo{author}{Paul, C.C.}, \bibinfo{year}{2014}.
\newblock \bibinfo{title}{Structural zeroes and zero-inflated models}.
\newblock \bibinfo{journal}{Shanghai archives of psychiatry}
  \bibinfo{volume}{26}, \bibinfo{pages}{236}.
\bibitem[{Hulchanski(2010)}]{hulchanski2010}
\bibinfo{author}{Hulchanski, J.D.}, \bibinfo{year}{2010}.
\newblock \bibinfo{title}{The three cities within toronto: Income polarization
  among toronto's neighbourhoods, 1970--2005. university of toronto}.
\bibitem[{Klein and Smart(2017)}]{klein2017}
\bibinfo{author}{Klein, N.J.}, \bibinfo{author}{Smart, M.J.},
  \bibinfo{year}{2017}.
\newblock \bibinfo{title}{Car today, gone tomorrow: The ephemeral car in
  low-income, immigrant and minority families}.
\newblock \bibinfo{journal}{Transportation} \bibinfo{volume}{44},
  \bibinfo{pages}{495--510}.
\bibitem[{Klein and Smart(2019)}]{klein2019}
\bibinfo{author}{Klein, N.J.}, \bibinfo{author}{Smart, M.J.},
  \bibinfo{year}{2019}.
\newblock \bibinfo{title}{Life events, poverty, and car ownership in the united
  states}.
\newblock \bibinfo{journal}{Journal of Transport and Land Use}
  \bibinfo{volume}{12}, \bibinfo{pages}{395--418}.
\bibitem[{Kramer(2018)}]{Kramer2018}
\bibinfo{author}{Kramer, A.}, \bibinfo{year}{2018}.
\newblock \bibinfo{title}{The unaffordable city: Housing and transit in north
  american cities}.
\newblock \bibinfo{journal}{Cities} \bibinfo{volume}{83},
  \bibinfo{pages}{1--10}.
\newblock \DOIprefix\doi{https://doi.org/10.1016/j.cities.2018.05.013}.
\bibitem[{Lo et~al.(2011)Lo, Shalaby and Alshalalfah}]{lo2011}
\bibinfo{author}{Lo, L.}, \bibinfo{author}{Shalaby, A.},
  \bibinfo{author}{Alshalalfah, B.}, \bibinfo{year}{2011}.
\newblock \bibinfo{title}{Relationship between immigrant settlement patterns
  and transit use in the greater toronto area}.
\newblock \bibinfo{journal}{Journal of Urban Planning and Development}
  \bibinfo{volume}{137}, \bibinfo{pages}{470--476}.
\bibitem[{Lucas(2012)}]{Lucas2012}
\bibinfo{author}{Lucas, K.}, \bibinfo{year}{2012}.
\newblock \bibinfo{title}{Transport and social exclusion: Where are we now?}
\newblock \bibinfo{journal}{Transport Policy} \bibinfo{volume}{20},
  \bibinfo{pages}{105--113}.
\newblock \DOIprefix\doi{https://doi.org/10.1016/j.tranpol.2012.01.013}.
  \bibinfo{note}{uRBAN TRANSPORT INITIATIVES}.
\bibitem[{Lucas et~al.(2009)Lucas, Tyler and Christodoulou}]{Lucas2009}
\bibinfo{author}{Lucas, K.}, \bibinfo{author}{Tyler, S.},
  \bibinfo{author}{Christodoulou, G.}, \bibinfo{year}{2009}.
\newblock \bibinfo{title}{Assessing the ‘value’ of new transport
  initiatives in deprived neighbourhoods in the uk}.
\newblock \bibinfo{journal}{Transport Policy} \bibinfo{volume}{16},
  \bibinfo{pages}{115--122}.
\newblock \DOIprefix\doi{https://doi.org/10.1016/j.tranpol.2009.02.004}.
\bibitem[{Martens(2016)}]{Martens2016}
\bibinfo{author}{Martens, K.}, \bibinfo{year}{2016}.
\newblock \bibinfo{title}{Transport justice: Designing fair transportation
  systems}.
\newblock \bibinfo{publisher}{Routledge}.
\bibitem[{Oswin(2014)}]{oswin2014}
\bibinfo{author}{Oswin, N.}, \bibinfo{year}{2014}.
\newblock \bibinfo{title}{Queer theory}, in: \bibinfo{booktitle}{The Routledge
  handbook of mobilities}. \bibinfo{publisher}{Routledge}, pp.
  \bibinfo{pages}{105--113}.
\bibitem[{Paez et~al.(2009)Paez, Ruben, Faber, Morency and Roorda}]{Paez2009}
\bibinfo{author}{Paez, A.}, \bibinfo{author}{Ruben, M.},
  \bibinfo{author}{Faber, S.}, \bibinfo{author}{Morency, C.},
  \bibinfo{author}{Roorda, M.}, \bibinfo{year}{2009}.
\newblock \bibinfo{title}{Mobility and social exclusion in canadian
  communities: An empirical investigation of canadian communities}.
\bibitem[{Palm et~al.(2020)Palm, Shalaby and Farber}]{Palm2020Social}
\bibinfo{author}{Palm, M.}, \bibinfo{author}{Shalaby, A.},
  \bibinfo{author}{Farber, S.}, \bibinfo{year}{2020}.
\newblock \bibinfo{title}{Social equity and bus on-time performance in
  canada’s largest city}.
\newblock \bibinfo{journal}{Transportation Research Record}
  \bibinfo{volume}{2674}, \bibinfo{pages}{329--342}.
\newblock \DOIprefix\doi{10.1177/0361198120944923}.
\bibitem[{Potoglou and Kanaroglou(2008)}]{Potoglou2008}
\bibinfo{author}{Potoglou, D.}, \bibinfo{author}{Kanaroglou, P.S.},
  \bibinfo{year}{2008}.
\newblock \bibinfo{title}{Modelling car ownership in urban areas: a case study
  of hamilton, canada}.
\newblock \bibinfo{journal}{Journal of Transport Geography}
  \bibinfo{volume}{16}, \bibinfo{pages}{42--54}.
\newblock \DOIprefix\doi{https://doi.org/10.1016/j.jtrangeo.2007.01.006}.
\bibitem[{Pucher(2002)}]{pucher2002}
\bibinfo{author}{Pucher, J.}, \bibinfo{year}{2002}.
\newblock \bibinfo{title}{Renaissance of public transport in the united
  states?}
\newblock \bibinfo{journal}{Transportation Quarterly} \bibinfo{volume}{56},
  \bibinfo{pages}{33--49}.
\bibitem[{Pucher and Renne(2003)}]{pucher2003}
\bibinfo{author}{Pucher, J.}, \bibinfo{author}{Renne, J.L.},
  \bibinfo{year}{2003}.
\newblock \bibinfo{title}{Socioeconomics of urban travel. evidence from the
  2001 nhts.} .
\bibitem[{Raphael and Rice(2002)}]{Raphael2002}
\bibinfo{author}{Raphael, S.}, \bibinfo{author}{Rice, L.},
  \bibinfo{year}{2002}.
\newblock \bibinfo{title}{Car ownership, employment, and earnings}.
\newblock \bibinfo{journal}{Journal of Urban Economics} \bibinfo{volume}{52},
  \bibinfo{pages}{109--130}.
\newblock \DOIprefix\doi{https://doi.org/10.1016/S0094-1190(02)00017-7}.
\bibitem[{Richmond(2001)}]{Richmond2001}
\bibinfo{author}{Richmond, J.}, \bibinfo{year}{2001}.
\newblock \bibinfo{title}{A whole-system approach to evaluating urban transit
  investments}.
\newblock \bibinfo{journal}{Transport Reviews} \bibinfo{volume}{21},
  \bibinfo{pages}{141--179}.
\newblock \DOIprefix\doi{10.1080/01441640116962}.
\bibitem[{Rosenbloom(1998)}]{rosenbloom1998}
\bibinfo{author}{Rosenbloom, S.}, \bibinfo{year}{1998}.
\newblock \bibinfo{title}{Transit markets of the future: the challenge of
  change}. volume~\bibinfo{volume}{28}.
\newblock \bibinfo{publisher}{Transportation Research Board}.
\bibitem[{Sanchez et~al.(2004)Sanchez, Shen and Peng}]{Sanchez2004}
\bibinfo{author}{Sanchez, T.W.}, \bibinfo{author}{Shen, Q.},
  \bibinfo{author}{Peng, Z.R.}, \bibinfo{year}{2004}.
\newblock \bibinfo{title}{Transit mobility, jobs access and low-income labour
  participation in us metropolitan areas}.
\newblock \bibinfo{journal}{Urban Studies} \bibinfo{volume}{41},
  \bibinfo{pages}{1313--1331}.
\newblock \DOIprefix\doi{10.1080/0042098042000214815}.
\bibitem[{Scholten and Joelsson(2019)}]{scholten2019}
\bibinfo{author}{Scholten, C.L.}, \bibinfo{author}{Joelsson, T.},
  \bibinfo{year}{2019}.
\newblock \bibinfo{title}{Integrating gender into transport planning: From one
  to many tracks}.
\newblock \bibinfo{publisher}{Springer}.
\bibitem[{Stanley et~al.(2011)Stanley, Hensher, Stanley, Currie, Greene and
  Vella-Brodrick}]{stanley2011}
\bibinfo{author}{Stanley, J.}, \bibinfo{author}{Hensher, D.A.},
  \bibinfo{author}{Stanley, J.}, \bibinfo{author}{Currie, G.},
  \bibinfo{author}{Greene, W.H.}, \bibinfo{author}{Vella-Brodrick, D.},
  \bibinfo{year}{2011}.
\newblock \bibinfo{title}{Social exclusion and the value of mobility}.
\newblock \bibinfo{journal}{Journal of Transport Economics and Policy (JTEP)}
  \bibinfo{volume}{45}, \bibinfo{pages}{197--222}.
\bibitem[{{Statistics Canada}(2021)}]{Statistics2021}
\bibinfo{author}{{Statistics Canada}}, \bibinfo{year}{2021}.
\newblock \bibinfo{title}{Annual demographic estimates: Subprovincial areas}.
\newblock
  \bibinfo{howpublished}{\url{https://www150.statcan.gc.ca/t1/tbl1/en/tv.action?pid=1710014201}}.
\newblock \bibinfo{note}{Accessed: 01.31.2021}.
\bibitem[{Staub and Winkelmann(2013)}]{Staub2013}
\bibinfo{author}{Staub, K.E.}, \bibinfo{author}{Winkelmann, R.},
  \bibinfo{year}{2013}.
\newblock \bibinfo{title}{Consistent estimation of zero-inflated count models}.
\newblock \bibinfo{journal}{Health Economics} \bibinfo{volume}{22},
  \bibinfo{pages}{673--686}.
\newblock \DOIprefix\doi{https://doi.org/10.1002/hec.2844}.
\bibitem[{Sultana et~al.(2018)Sultana, Mishra, Cherry, Golias and {Tabrizizadeh
  Jeffers}}]{Sultana2018}
\bibinfo{author}{Sultana, Z.}, \bibinfo{author}{Mishra, S.},
  \bibinfo{author}{Cherry, C.R.}, \bibinfo{author}{Golias, M.M.},
  \bibinfo{author}{{Tabrizizadeh Jeffers}, S.}, \bibinfo{year}{2018}.
\newblock \bibinfo{title}{Modeling frequency of rural demand response transit
  trips}.
\newblock \bibinfo{journal}{Transportation Research Part A: Policy and
  Practice} \bibinfo{volume}{118}, \bibinfo{pages}{494--505}.
\newblock \DOIprefix\doi{https://doi.org/10.1016/j.tra.2018.10.006}.
\bibitem[{Train(2009)}]{train2009}
\bibinfo{author}{Train, K.E.}, \bibinfo{year}{2009}.
\newblock \bibinfo{title}{Discrete Choice Methods with Simulation}.
\newblock \bibinfo{edition}{2} ed., \bibinfo{publisher}{Cambridge University
  Press}.
\newblock \DOIprefix\doi{10.1017/CBO9780511805271}.
\bibitem[{Tyndall(2017)}]{Tyndall2017}
\bibinfo{author}{Tyndall, J.}, \bibinfo{year}{2017}.
\newblock \bibinfo{title}{Waiting for the r train: Public transportation and
  employment}.
\newblock \bibinfo{journal}{Urban Studies} \bibinfo{volume}{54},
  \bibinfo{pages}{520--537}.
\newblock \DOIprefix\doi{10.1177/0042098015594079}.
\bibitem[{Walks(2018)}]{Walks2018}
\bibinfo{author}{Walks, A.}, \bibinfo{year}{2018}.
\newblock \bibinfo{title}{Driving the poor into debt? automobile loans,
  transport disadvantage, and automobile dependence}.
\newblock \bibinfo{journal}{Transport Policy} \bibinfo{volume}{65},
  \bibinfo{pages}{137--149}.
\newblock \DOIprefix\doi{https://doi.org/10.1016/j.tranpol.2017.01.001}.
  \bibinfo{note}{household transport costs, economic stress and energy
  vulnerability}.

\end{thebibliography}

\section{Acknowledgements}
We are grateful to the University of Toronto Data Management Group who generously provided access to TTS data via their remote desktop servers.

\newpage
\appendix
\section{New transit trips after transit improvement} \label{sec:new_transit_trip}
\begin{figure}[!ht]
    \centering
    \includegraphics[width=0.7\textwidth]{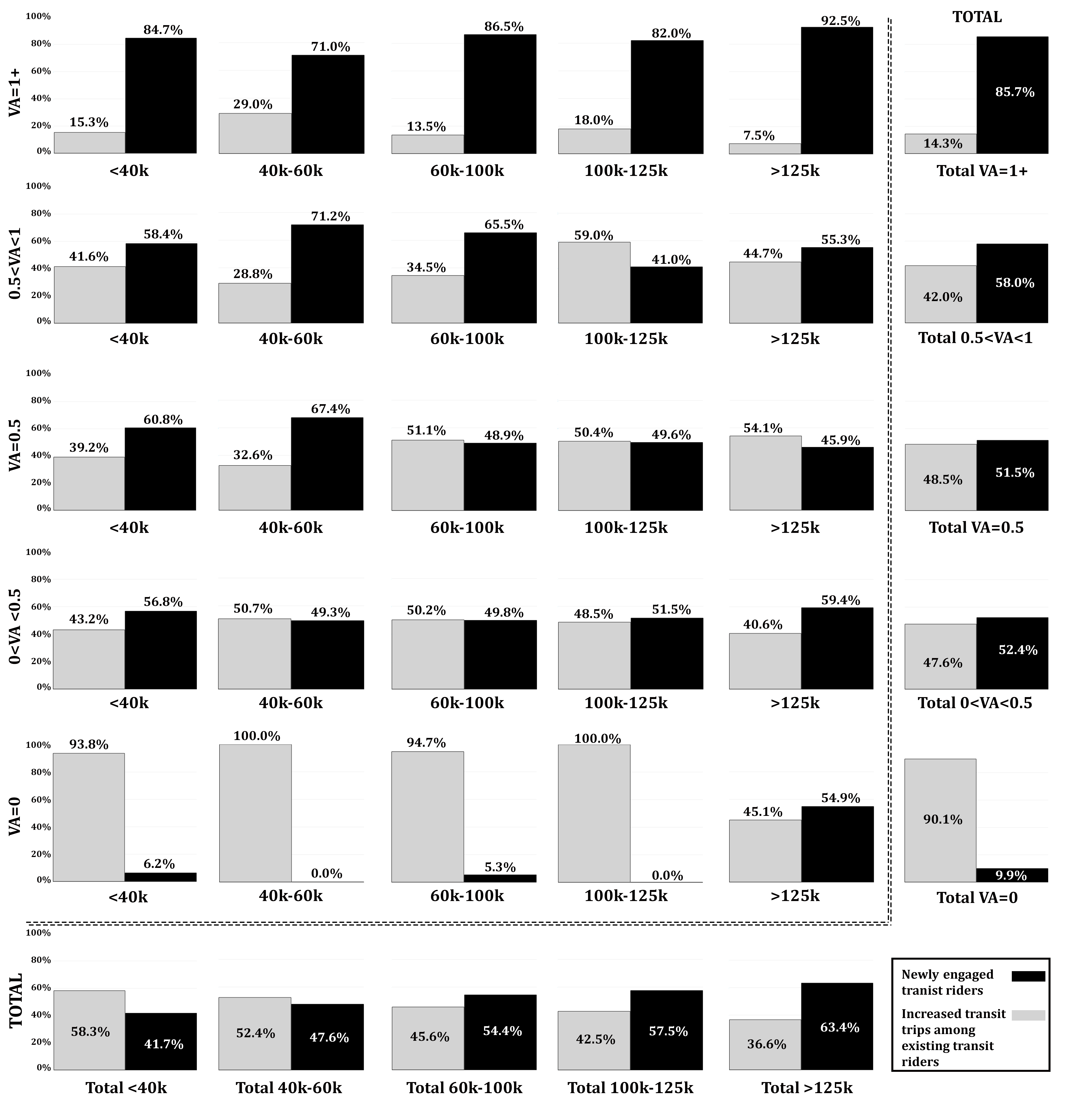}
    \caption{The ratio of newly generated transit trips given existing and new users after transit improvement (200k)}
    \label{fig:newly_generated_tranist_trips}
\end{figure}

Figure~\ref{fig:newly_generated_tranist_trips} shows what percentage of new transit trips in each stratum originates from existing transit users and what percentage from new transit riders. It indicates improving transit accessibility has a significant impact on non-riders of households owning one or more cars per adult — i.e., VA = 1+. These individuals have the highest sensitivity to transit accessibility improvement. On the other hand, most zero-car families with the least elasticity (Figure~\ref{fig:heatmap_elasticity_estimates}) are already transit riders. Therefore, accessibility improvements entice more existing riders of carless households to take more transit trips than non-riders.

Among carless households, non-riders of wealthy families are easily absorbed after transit accessibility improvement. However, the majority of low-income individuals are already transit riders. Therefore, they tend to increase their existing transit trips after accessibility gains. It illustrates that this stratum is still an unsaturated market, and its individuals need to be provided with transit investment. Interestingly, three car-deficit groups have almost equal potential for whether generating new transit trips or expanding their current transit trips.

\end{document}